\documentclass[twocolumn,twocolappendix]{aastex631}

\usepackage{showyourwork}

\usepackage{siunitx}
\usepackage{blindtext}          
\usepackage{xcolor}
\usepackage{amssymb}

\usepackage{datatool, filecontents}
\DTLsetseparator{,}
\DTLloaddb[noheader, keys={thekey,thevalue}]{variables}{variables.dat}
\newcommand{\var}[1]{\DTLfetch{variables}{thekey}{#1}{thevalue}}

\usepackage{comment}						 
\includecomment{comment}
\specialcomment{note}{\begingroup\sffamily\color{red!40!green!70!blue!90}}{\endgroup}

\usepackage[disable]{todonotes}			
\usepackage{regexpatch}
\makeatletter
\xpatchcmd{\@todo}{\setkeys{todonotes}{#1}}{\setkeys{todonotes}{inline,#1}}{}{}
\makeatother

\newcommand{\rev}[1]{{#1}}
\newcommand{\revii}[1]{{#1}}
\newcommand{\reviii}[1]{{#1}}

\newcommand{\windowsize}{25}
\newcommand{\minMagnitude}{16}         

\newcommand{\prSmin}{10}
\newcommand{\prSmax}{1000}
\newcommand{\prWRRmin}{10^{-5}}
\newcommand{\prWRRmax}{0.1}
\newcommand{\prRmin}{0.1}
\newcommand{\prRmax}{15}

\newcommand{\frgh}{0.8}
\newcommand{\sigmaR}{\SI{2}{\percent}}
\newcommand{\sigmaS}{\SI{5}{\percent}}
\newcommand{\avgRadiusChange}{\SI{15}{\percent}}

\DeclareSIUnit\mSun{M_\odot}
\DeclareSIUnit\Msun{M_\odot}
\DeclareSIUnit\mStar{M_\star}
\DeclareSIUnit\Mstar{M_\star}
\DeclareSIUnit\mEarth{M_\oplus}
\DeclareSIUnit\Mearth{M_\oplus}
\DeclareSIUnit\rEarth{R_\oplus}
\DeclareSIUnit\Rearth{R_\oplus}
\DeclareSIUnit\year{yr}
\DeclareSIUnit\au{au}
\DeclareSIUnit\dex{dex}
\DeclareSIUnit\ppm{ppm}
\DeclareSIUnit\eV{eV}
\DeclareSIUnit\parsec{pc}

\newcommand{\code}[1]{\texttt{#1}}
\newcommand{\project}[1]{\textsl{#1}}

\newcommand{\bioverse}{\code{Bioverse}}

\newcommand{\rst}{\project{Nancy Grace Roman Space Telescope}}
\newcommand{\plato}{\project{PLATO}}
\newcommand{\cheops}{\project{CHEOPS}}
\newcommand{\kepler}{\project{Kepler}}
\newcommand{\ktwo}{\project{K2}}
\newcommand{\tess}{\project{TESS}}
\newcommand{\ariel}{\project{Ariel}}
\newcommand{\nautilus}{\project{Nautilus}}
\newcommand{\life}{\project{LIFE}}
\newcommand{\hwo}{\project{Habitable Worlds Observatory}}
\newcommand{\gclef}{\project{G-CLEF}}
\newcommand{\gmt}{\project{Giant Magellan Telescope}}
\newcommand{\andes}{\project{ANDES}}
\newcommand{\elt}{\project{European Extremely Large Telescope}}
\newcommand{\modhis}{\project{MODHIS}}
\newcommand{\tmt}{\project{Thirty Meter Telescope}}

\newcommand{\gaia}{\textsl{Gaia}}

\begin{document}
\begin{filecontents*}{variables.dat}
N_optimistic,500
percentage_transiting,\SI{1.5}{\percent}
platoMinfrgh,0.1
platoMinfrghHundred,0.2
platoMinfrghHundred_rho,0.1
retrSthresh,\SI{276}{\watt\per\meter\squared}
injectedSthresh,\SI{280}{\watt\per\meter\squared}
N_stars_FGK,15702
N_stars_M,34371
N_FGK,40
N_M,228
N_FGK_err,6
N_M_err,14
N_plato,300
wrr,0.005
\end{filecontents*}

\title{Bioverse: The Habitable Zone Inner Edge Discontinuity as an Imprint of Runaway Greenhouse Climates on Exoplanet Demographics}

\author[0000-0001-8355-2107]{Martin Schlecker}
\affiliation{Steward Observatory, The University of Arizona, Tucson, AZ 85721, USA; \href{mailto:schlecker@arizona.edu}{schlecker@arizona.edu}}
\author[0000-0003-3714-5855]{D\'{a}niel Apai}
\affiliation{Steward Observatory, The University of Arizona, Tucson, AZ 85721, USA; \href{mailto:schlecker@arizona.edu}{schlecker@arizona.edu}}
\affiliation{Lunar and Planetary Laboratory, The University of Arizona, Tucson, AZ 85721, USA}
\author[0000-0002-3286-7683]{Tim Lichtenberg}
\affiliation{Kapteyn Astronomical Institute, University of Groningen, PO Box 800, 9700 AV Groningen, The Netherlands}
\author[0000-0003-4500-8850]{Galen Bergsten}
\affiliation{Lunar and Planetary Laboratory, The University of Arizona, Tucson, AZ 85721, USA}
\author[0000-0001-8106-6164]{Arnaud Salvador}
\affiliation{Lunar and Planetary Laboratory, The University of Arizona, Tucson, AZ 85721, USA}
\author[0000-0003-3702-0382]{Kevin K.\ Hardegree-Ullman}
\affiliation{Steward Observatory, The University of Arizona, Tucson, AZ 85721, USA; \href{mailto:schlecker@arizona.edu}{schlecker@arizona.edu}}

\begin{abstract}
Long-term magma ocean phases on rocky exoplanets orbiting closer to their star than the runaway greenhouse threshold -- the inner edge of the classical habitable zone -- may offer insights into the physical and chemical processes that distinguish potentially habitable worlds from others.
Thermal stratification of runaway planets is expected to significantly inflate their atmospheres, potentially providing observational access to the runaway greenhouse transition in the form of a ``habitable zone inner edge discontinuity'' in radius--density space.
Here, we use \bioverse, a statistical framework combining contextual information from the overall planet population with a survey simulator, to assess the ability of ground- and space-based telescopes to test this hypothesis.

We find that the demographic imprint of the runaway greenhouse transition is likely detectable with high-precision transit photometry for sample sizes $\gtrsim$~100~planets if at least \SI{\sim 10}{\percent} of those orbiting closer than the habitable zone inner edge harbor runaway climates.
Our survey simulations suggest that in the near future, ESA's \plato\ mission will be the most promising survey to probe the habitable zone inner edge discontinuity.
We determine survey strategies that maximize the diagnostic power of the obtained data and identify as key mission design drivers: 1. A follow-up campaign of planetary mass measurements and 2. The fraction of low-mass stars in the target sample.
Observational constraints on the runaway greenhouse transition will provide crucial insights into the distribution of atmospheric volatiles among rocky exoplanets, which may help to identify the nearest potentially habitable worlds.
\end{abstract}

\section{Introduction}
Despite recent advancements in observational techniques, our understanding of terrestrial-sized planets remains woefully limited, with fundamental aspects of their nature, composition, and potential habitability still largely unknown.
Due to the inherent biases of current exoplanet detection techniques, the best-studied category of rocky exoplanets at present is that of hot or warm, close-in planets~\citep{2019AREPS..47..141J,2021JGRE..12606639B}.
These experience thermal states that are in some aspects comparable to the ones of the inner solar system bodies at early stages of their evolution~\citep{Ikoma2018,2021ChEG...81l5735C}, which likely profoundly affected the distribution of volatiles between planetary core, mantle, and atmosphere.
Studying the geophysical state of hot exoplanets can thus inform our understanding of the early evolutionary stages of Earth and other habitable worlds~\citep{Lichtenberg2022,Krijt2022}.

An example from the solar system for the potential significance of these early stages are the divergent atmospheric evolutions of Venus and Earth~\citep[e.g.,][]{2019JGRE..124.2015K,2021JGRE..12606643K,Salvador2023b}.
While having accreted from a similar mass reservoir~\citep{2020plas.book..287R,2020SSRv..216...55K,2020SSRv..216...27M,2020plas.book....3Z} and despite their similar bulk properties~\citep{Smrekar2018}, they evolved into planets with very different surface conditions~\citep{1982Sci...216..630D,Kasting1988,Hamano2013,Kane2014,Way2020,Turbet2021}.
Both planets likely underwent a giant impact phase~\citep{2020plas.book..287R,2020NatGe..13..265G,Liu2022} that melted their mantles~\citep{2012AREPS..40..113E,2018RSPTA.37680109S,Lichtenberg2022}.
Magma ocean states play a substantial role in establishing the long-term geophysical and climatic regimes of rocky planets~\citep{2020ChEG...80l5594F}, in particular owing to efficient heat and volatile transfers between interior and atmosphere in the absence of a stiff boundary separating them~\citep{2021ApJ...909L..22K,Dorn2021,Salvador2023b}.
Due to these similar formation sequences, it was commonly assumed that the divergence of Venus and Earth -- in particular Venus' water loss -- occurred late in their evolution~\citep[e.g.,][]{Way2020}.

Yet, \citet{Hamano2013} suggested that the present-day dry conditions on Venus may have been directly inherited from the early magma ocean stage. 
If a strongly infrared-absorbing, condensable species such as water was dominant in the atmosphere, the resulting strong thermal blanketing effect would prevent the planet from efficiently radiating to space and maintaining the surface molten~\citep{Ingersoll1969,Kasting1988,2010ppc..book.....P,Goldblatt2013,2015ExA....40..449L,Salvador2017}.
This runaway greenhouse state can extend the magma ocean stage to hundreds of \SI{}{\mega\year}~\citep{2016ApJ...829...63S,2021AsBio..21.1325B}, enough to remove the entire water reservoir from a rocky planet by H$_2$O photolysis and subsequent hydrodynamic escape of hydrogen~\citep{2013ApJ...778..154W,2014ApJ...785L..20W,Luger2015}.
For Venus, \rev{the atmospheric composition~\citep{2020NatGe..13..265G} and comprehensive analysis of meteoritic samples across the Solar System~\citep{2018SSRv..214...36A,2022Natur.611..245B}} suggest that it went through this phase.
Although the past presence or absence of a Venusian water ocean has not been definitely established~\citep{Raymond2006,Raymond2007,Hamano2013,Way2016,2019JGRE..124.2015K,2021JGRE..12606643K,Turbet2021,2023PNAS..12009751W}, a transient habitable phase cannot be conclusively ruled out~\citep[e.g.,][]{Way2016,Salvador2017,Krissansen-Totton2021}.

The runaway greenhouse transition is a robust prediction from climate models~\citep{Kasting1988,Nakajima1992,Goldblatt2012,Forget2014,Boukrouche2021,2022A&A...658A..40C}, and its impact on planetary bulk properties has been shown to be in the detectable range of current astronomical instrumentation~\citep{Goldblatt2015}.
In particular, planets in a runaway greenhouse state are expected to be thermally inflated~\citep{Turbet2019,Turbet2020,Mousis2020}, which directly increases their transit radii by an amount that is a function of the water content.
However, dissolution of water~\citep[e.g.,][]{Elkins-Tanton2008,Hier-Majumder2017,Salvador2023} in the magma may decrease this effect~\citep{Dorn2021}, and chemical exchange between core and mantle material may influence the amount and speciation of outgassed volatiles that are visible in the atmosphere via transmission spectroscopy~\citep{2021ApJ...914L...4L,Schlichting2022}.
Astronomical observations of planets that are currently in a runaway greenhouse state may thus constrain properties of their mantles and establish an observational connection between exoplanetary interiors and atmospheres~\citep{Lichtenberg2022,Wordsworth2022}.

Of particular relevance is that the radiation-induced transition between a runaway greenhouse state and a temperate climate is thought to occur at a relatively sharp instellation threshold~\citep{Goldblatt2013,Leconte2013,Kopparapu2013}.
Consequently, the instellation at which the runaway greenhouse transition occurs is aptly considered to be the inner boundary of the habitable zone~\citep[e.g.,][]{Ramirez2018,Salvador2023b}.
Its prevalent definition refers to the possibility of a sustained liquid water body on the surface of an Earth-like planet with an oxidized CO$_2$/H$_2$O/N$_2$-rich atmosphere~\citep[][]{Kasting1993,Kopparapu2013,Kopparapu2014};
its exact spatial location and extent may be strongly influenced by the interior and atmosphere oxidation state and resulting atmosphere composition~\citep{2011ApJ...734L..13P,2017ApJ...837L...4R,2018ApJ...858...72R,2019ApJ...875...31K,2020ApJ...896..115G,2022JGRE..12707456G,2023ApJ...942L..20H}.
The fundamental concept of a habitable zone dates back centuries~\citep{Newton1687,Whewell1858,Shapley1953,Huang1959}, and its modern form has proven popular in the planetary literature\footnote{At the time of writing, a search of the term ``habitable zone'' in the titles and abstracts of refereed articles in the National Aeronautics and Space Administration (NASA) Astrophysics Data System returned $\sim\num{1700}$ results.}.
However, it should be emphasized that the habitable zone, as it stands today, is merely a concept based on theoretical predictions and geochemical evidence from one planet -- Earth~\citep{2020SciA....6.1420C} -- and its general validity remains controversial~\citep[e.g.,][]{Cockell2016,Moore2017,Tuchow2023}.
The question naturally emerges if a planetary habitable zone -- in its common form with boundaries defined by stellar irradiation -- is a predictive theoretical concept and how the habitable zone hypothesis can be tested observationally.

Observational tests being considered include searches for direct evidence of liquid water conveyed by ocean glint~\citep{Williams2008,Robinson2010,Lustig-Yaeger2018} or water vapor in planetary atmospheres~\citep{Suissa2020}.
A different approach relies on comparative planetology:
aiming for a statistical detection in a planet population provides robustness against ambiguity that could otherwise arise from individual variations in a planet's composition or geophysical history~\citep{Checlair2019,Apai2019a}.
Tests suggested in the literature include determining, for a range of orbital distances, atmospheric H$_{2}$O and CO$_2$ abundances, planetary albedos~\citep{Bean2017,Bixel2021}, or colors~\citep{Crow2011,Bixel2020}, or testing the relationship between CO$_2$ partial pressure and incident flux~\citep{Lehmer2020}.
All these tests require surveying a large enough sample of terrestrial-sized planets with next-generation instruments, rendering them out of reach in the immediate future.

Here, we explore the feasibility of a statistical test of the habitable zone hypothesis by surveying planetary bulk properties close to its inner edge, the runaway greenhouse transition.
Our goal is to assess the ability of near-future transit surveys to test the hypothesis that the runaway greenhouse effect causes a discontinuity of planetary radii and bulk densities when ordered by receiving instellation~\citep{Turbet2019}.
Our main tool for this is \bioverse, a simulation framework for assessing the statistical power of exoplanet surveys~\citep{Bixel2021}.
It consists of a sample generator that populates stars from the \gaia\ catalog~\citep[][]{Hardegree-Ullman2023,Gaia_mission,Gaia_DR3} with planetary systems based on state-of-the-art occurrence rates~\citep{Bergsten2022}, a flexible survey simulator that allows for a broad range of trade studies, and a hypothesis testing module that quantifies the survey's ability to detect a previously injected trend.
Trying different instrumentation and survey designs, we use \bioverse\ to recover runaway greenhouse-induced effects based on model predictions~\citep{Turbet2020,Dorn2021} that we inject into a baseline planet population.

In particular, we test the capability of the PLAnetary Transits and Oscillation of stars~\citep[\plato, ][]{Rauer2016} mission, which will measure the radii of a large number of terrestrial-sized planets, to detect the radius/density discontinuity and determine its sensitivity to model assumptions and fundamental processes.
We then perform a parameter study to explore which trades in the survey design of a \plato-like mission maximize its diagnostic power to test runaway greenhouse climate models through the detection of the habitable zone inner edge discontinuity.

We organize the paper as follows:
Section~\ref{sec:met_baseline} introduces the baseline model we use to produce synthetic star and planet samples.
In Section~\ref{sec:met_rghmodel}, we describe the model component that produces runaway greenhouse-induced transit radius changes.
Section~\ref{sec:met_surveys-hypotests} explains our survey simulations and hypothesis tests.
We present our results in Section~\ref{sec:results} before interpreting them in Section~\ref{sec:discussion}.
Finally, we summarize our findings in Section~\ref{sec:conclusions}.

\section{Baseline Model}\label{sec:met_baseline}
\begin{figure*}
    \begin{centering}
        \includegraphics[width=\hsize]{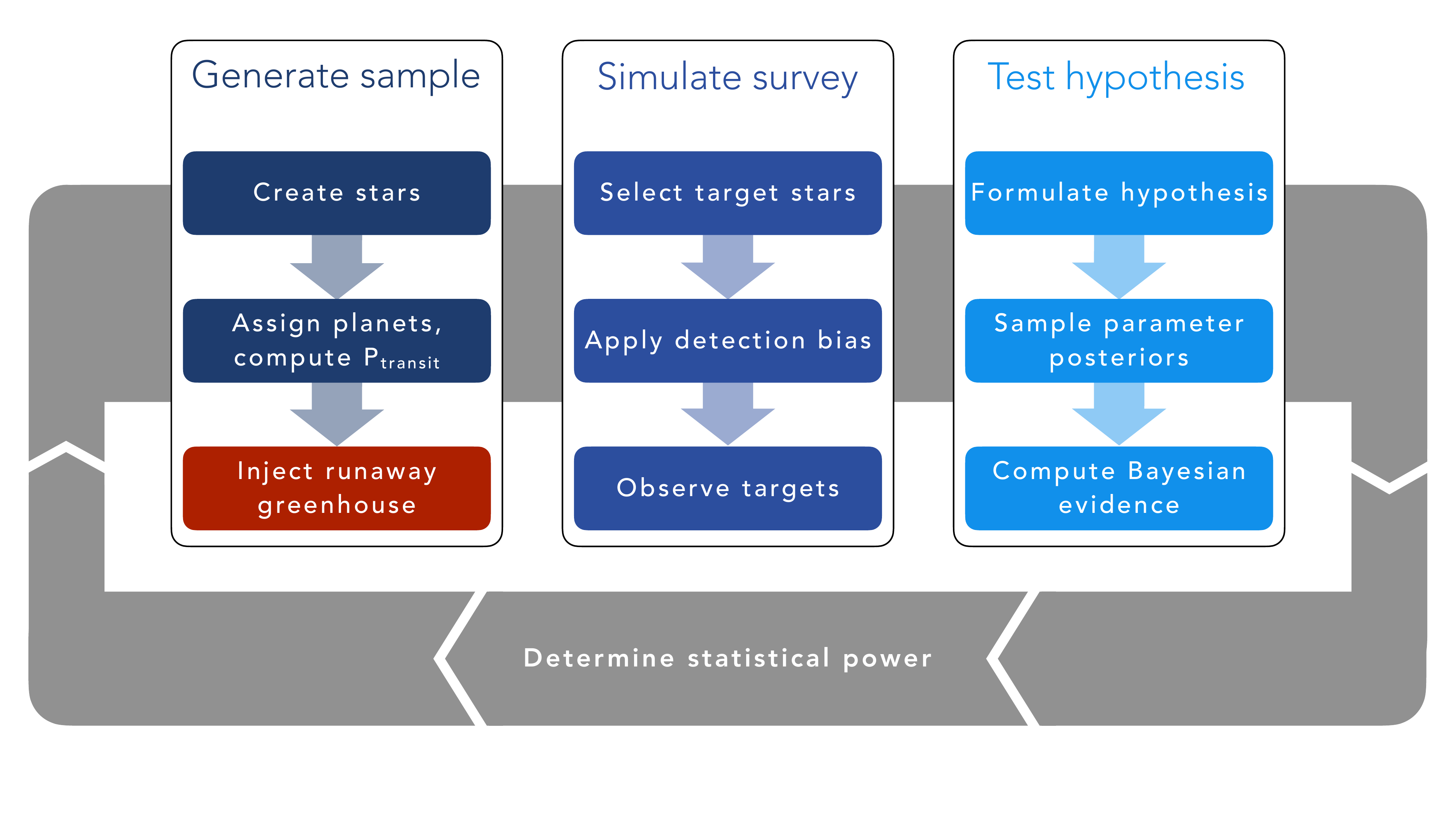}
        \caption{Workflow of our hypothesis testing with \bioverse. First, we generate a sample of stars and populate them with planets based on \kepler\ demographics.
            A fraction of them are then assigned a runaway greenhouse climate based on the model described in Section~\ref{sec:met_rghmodel}.
        We then simulate an exoplanet survey, whereby selection effects and detection biases are introduced. Finally, we test the runaway greenhouse hypothesis based on data from the survey simulation.
        By iterating through these steps, we compute the statistical power of testing the hypothesis for different survey designs.}
        \label{fig:flowchart}
    \end{centering}
\end{figure*}

The goal of this study is to determine -- for different configurations of near-future exoplanet surveys -- the confidence level with which the runaway greenhouse threshold can be detected statistically. 
Our basic methodology was as follows:
We expanded the \bioverse\ framework~\citep{Bixel2020,Bixel2021}\footnote{\bioverse\ is actively maintained and documented open source software written in Python. Its latest version and documentation can be found at \url{https://github.com/danielapai/bioverse}.} to generate synthetic samples of stars that host planets according to the observed exoplanet demographics.
We then adapted planetary bulk properties as predicted from models of runaway greenhouse atmospheres, simulated observations of the planets, and computed Bayesian evidences in favor of a habitable zone inner edge discontinuity (see diagram in Figure~\ref{fig:flowchart}).
In this section, we review the source of the stellar sample, the modeled luminosity evolution, the generation of a synthetic planet sample, and the orbital parameters of the planets.
An overview of our key assumptions and model parameters can be found in Table~\ref{tab:params_table}.

\begin{deluxetable*}{lrll}
\tablecaption{Key assumptions and model parameters used in our simulation setup}
\tablehead{\colhead{Parameter} & \colhead{Value} & \colhead{Unit} & \colhead{Description}}
\startdata
\textbf{Stellar sample} &  &  &  \\
$G_\mathrm{max}$ & 16 &  & Maximum Gaia magnitude \\
$M_\mathrm{\star, max}$ & 1.5 & $M_\odot$ & Maximum stellar mass \\
Luminosity evolution &  &  & \citet{Baraffe1998} \\
~\\ \textbf{Planetary parameters} &  &  &  \\
$M_\mathrm{P}$ & 0.1 -- 2.0 & $M_\oplus$ & Planetary mass range \\
$R_\mathrm{P, min}$ & 0.75 & $R_\oplus$ & Minimum planet radius \\
Baseline mass–radius relation &  &  & \citet{Zeng2016} \SI{100}{\percent} $\mathrm{MgSiO_3}$\tablenotemark{a} \\
$\delta_\mathrm{min}$ & 80 & ppm & Minimum transit depth \\
$P_\mathrm{max}$ & 500 & day & Maximum orbital period [day] \\
$S$ & 10 -- 2000 & \SI{}{\watt\per\meter\squared} & Net instellation range \\
$S_\mathrm{thresh}$ & 280 & \SI{}{\watt\per\meter\squared} & Threshold instellation for runaway greenhouse \\
~\\ \textbf{Runaway greenhouse model} &  &  &  \\
Runaway greenhouse atmospheric models &  &  & \citet{Turbet2020,Dorn2021} \\
$x_{H_2O}$ & \SIrange{e-5}{0.1}{} &  & \rev{Bulk water} mass fraction (fiducial case: 0.005) \\
$f_\mathrm{rgh}$ & \SIrange{0}{1}{} &  & Dilution factor (fiducial case: 0.8) \\
~\\ \textbf{Priors} &  &  &  \\
$\Pi(S_\mathrm{thresh}$) & [10, 1000] & \SI{}{\watt\per\meter\squared} & Uniform \\
$\Pi(x_{H_2O})$ & [\SI{e-5}{}, \SI{0.1}{}] &  & Log-uniform \\
$\Pi(f_\mathrm{rgh})$ & [0, 1] &  & Uniform \\
$\Pi(\langle R_\mathrm{P}\rangle_\mathrm{out})$ & [0, 15] & $R_\oplus$ & Mean radius of non-runaway planets, uniform
\enddata
\tablenotetext{a}{For a comparison with alternative interior compositions, see Appendix~\ref{app:MR_relation}.}
\label{tab:params_table}
\end{deluxetable*}

\subsection{Stellar sample from \gaia\ DR3}
The original \bioverse\ stellar catalog was generated randomly from the \citet{Chabrier2003} stellar mass function.
Improved parallax and photometric data from the \gaia\ mission made it possible to generate a homogeneous and complete stellar catalog out to about 100~pc, which became the new standard stellar catalog for \bioverse ~\citep{Hardegree-Ullman2023}.
Here, we briefly describe how we derived the stellar effective temperature $T_{\mathrm{eff}}$, luminosity $L_{\star}$, stellar radius $R_{\star}$, and stellar mass $M_{\star}$.

\citet{Hardegree-Ullman2023} used the \gaia\ Catalogue of Nearby Stars~\citep[hereafter GCNS,][]{Smart2021} as the basis for deriving stellar parameters for the \bioverse\ catalog.
The GCNS identified stars out to 120~pc and includes \gaia\ DR3 parallaxes and photometry in $G$, $G_{BP}$, and $G_{RP}$ bands~\citep{GaiaCollaboration2022} and $K_S$ band photometry from 2MASS~\citep{Cutri2003}, all of which were used in stellar classification.
From this information, we computed colors and absolute magnitudes and applied initial color-magnitude cuts to remove non-main sequence stars.
We derived effective temperatures primarily from color-temperature relations derived using the main sequence stellar parameters table from \citet{Pecaut2013}.
Luminosities were computed from absolute $G$-band magnitudes and a derived bolometric correction.
We computed stellar radii with the effective temperatures and luminosities using the Stefan-Boltzmann law or using absolute $K_S$-band magnitudes and an empirical radius-luminosity relation from \citet{Mann2015} for targets within the absolute magnitude range of M~dwarfs.
Finally, we derived masses from the mass-luminosity relation of \citet{Torres2010} for stars with $M_{\star}\gtrsim 0.7\,M_{\odot}$, and from that of \citet{Mann2019} for targets within the absolute magnitude range of M dwarfs.
The derived stellar \rev{parameters were} compared to \rev{measured parameters} for all known exoplanet hosts from the literature and were found to be consistent within 1\%, 3\%, and 5.5\% for $T_{\mathrm{eff}}$, $R_{\star}$, and $M_{\star}$, respectively, which are all below the typical measurement uncertainties of 3.3\%, 6.8\%, and 7.9\%, respectively~\rev{\citep{Hardegree-Ullman2023}}.
From this catalog, \bioverse\ samples stars within an isotropic distance from the solar system as required by the planetary sample size.

\subsection{Stellar luminosity evolution}
Planetary systems are hosted by stars of a wide range of ages, and stellar luminosities evolve with time.
Since the occurrence of a runaway greenhouse state is highly dependent on the amount of radiation received by the planet, and thus on the luminosity of the host star, we assigned age-dependent luminosities to our synthetic stars.

\begin{figure}[ht!]
    \script{plot_luminosity_tracks.py}
    \begin{centering}
        \includegraphics[width=\linewidth]{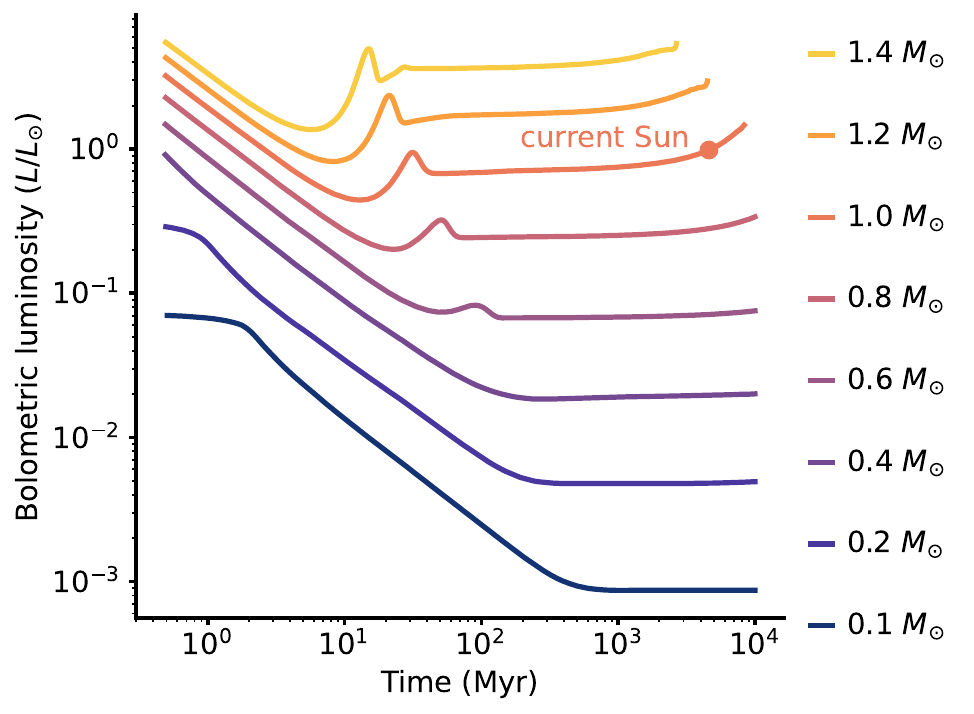}
        \caption{
            Bolometric luminosity tracks of stars with different masses, computed from stellar evolution models of \citet{Baraffe1998}.
            Low-mass stars, which make up the majority of stars in the solar neighborhood, undergo an extended early phase of several magnitudes higher luminosity before entering a lifetime of relative faintness.
        }
        \label{fig:luminosity_tracks}
    \end{centering}
\end{figure}

While stellar ages are notoriously poorly constrained~\cite[e.g.,][]{Adams2005}, the age distribution of planet host stars in the Solar neighborhood was shown to be broadly consistent with uniform~\citep{Reid2007,Gaidos2023}.
For our synthetic stars, we thus drew random ages from a uniform distribution from \SI{0}{\giga\year} to \SI{10}{\giga\year}.
We then assigned each star a luminosity from the mass-dependent evolutionary models of \citet{Baraffe1998}.
Figure~\ref{fig:luminosity_tracks} shows the corresponding luminosity evolution as a function of stellar mass and age.

\subsection{Synthetic planet sample}\label{sec:syn_planets}
Next, we assigned to the stellar sample planetary systems with frequencies, orbital parameters, and bulk properties derived from the \kepler\ mission.
We adopted the model from~\citet{Bergsten2022}, which defines the occurrence rate of small planets in radius and orbital period.
Following \citet{Youdin2011a}, their inferred occurrence rate density can be expressed in the form
\begin{equation}
    \frac{\mathrm{d}^2n}{\mathrm{d}R \, \partial P} = F_0 C_n g(R, P, M_\star),
\end{equation}
where $F_0$ represents the average number of planets per star, $C_n$ is a normalization constant, and the shape function $g(R, P, M_\star)$ describes the distribution of planets in radius, orbital period, and stellar host mass.
\bioverse\ generates planets based on the above occurrence rate density and assigns them to the previously generated stars.

\subsection{Orbit parameters and planet masses}\label{sec:met-orbits_masses}
Eccentric orbits alter the probability of a planet to transit~\citep[e.g.,][]{Barnes2007a}.
The distribution of eccentricities $e$ of exoplanets has been found to resemble a Beta function~\citep{Kipping2013b}, which we chose to draw synthetic eccentricities from.
Following~\citet{Kipping2013b}, we used a Beta distribution with parameters $a=0.867$ and $b=3.03$, and truncated the distribution at $e = 0.8$.
Assuming isotropic alignments of orbits, we assigned each planet an inclination drawn from a distribution uniform in $\cos(i)$.

To assign masses to our planets, we use the semi-empirical mass–radius relationship assuming a pure $\mathrm{MgSiO_3}$ composition from \citet{Zeng2016} (see green line in Figure~\ref{fig:radiusevolution}).
This represents the baseline bulk density before any climate-related effects are applied.

\subsection{Transit probability}
We model the occurrence of transits by assuming isotropic orientations of planetary orbits and calculating the impact parameters $b = a \cos(i)/R_\star$.
Following the approach in \citet{Bixel2021}, we further consider only planets with $|b| < 1$.
For these cases we calculate the transit depth
\begin{equation}\label{eq:transitdepth}
    \delta = \left( \frac{R_\mathrm{P}}{R_\star} \right)^2,
\end{equation}
which is relevant for the detection probability of the respective planet~(see Section~\ref{sec:sensitivity}).
Excluding all non-transiting planets diminishes the sample to \var{percentage_transiting} of its original size.

\section{Runaway Greenhouse Model}\label{sec:met_rghmodel}
The climate state of a planet has a direct influence on its apparent size measured by transit photometry~\citep{Turbet2019,Turbet2020,Mousis2020,2021ApJ...914...84A}.
With even a fraction of the Earth's water inventory, a planet absorbing more flux than the radiation limit of steam atmospheres \revii{was found to} enter a runaway greenhouse state resulting in a global magma ocean~\revii{\citep[][but see \citet{Selsis2023} for a contrasting viewpoint]{Lichtenberg2021c,Boukrouche2021}}.
We use predictions on transit atmospheric thickness from geophysical models to derive the change in transit radius and bulk density that planets with instellation-induced runaway greenhouse climates experience, depending on the distribution of water between planetary interior and atmosphere, and on the resulting thermal atmospheric structure~\citep{Dorn2021,Salvador2023}.

The net absorbed stellar \rev{fluxes of planets are a function of intrinsic atmospheric properties such as their albedo, which are} generally poorly constrained for planets outside the solar system~\citep[e.g.,][]{Angerhausen2015,Parmentier2018a,Mansfield2019}.
\rev{For instance, a planet with a high albedo may sustain temperate conditions closer to the star than the same planet with a lower albedo and located farther away from the star.}
Here, we assume global redistribution of incoming flux and a fixed Bond albedo of $0.3$, comparable to Earth's~\citep{Haar1971}.
We do not take into account additional heating sources such as tidal effects~\citep[e.g.,][]{Barnes2013}.

\begin{figure}
    \script{radiusevolution.py}
    \begin{centering}
        \includegraphics[width=\hsize]{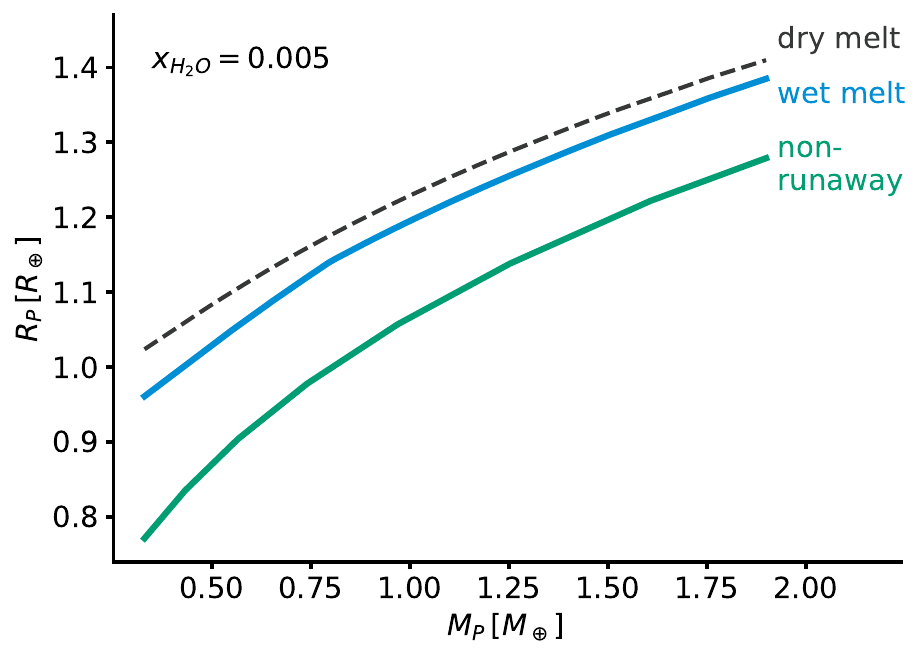}
    \end{centering}
    \begin{centering}
        \includegraphics[width=\hsize]{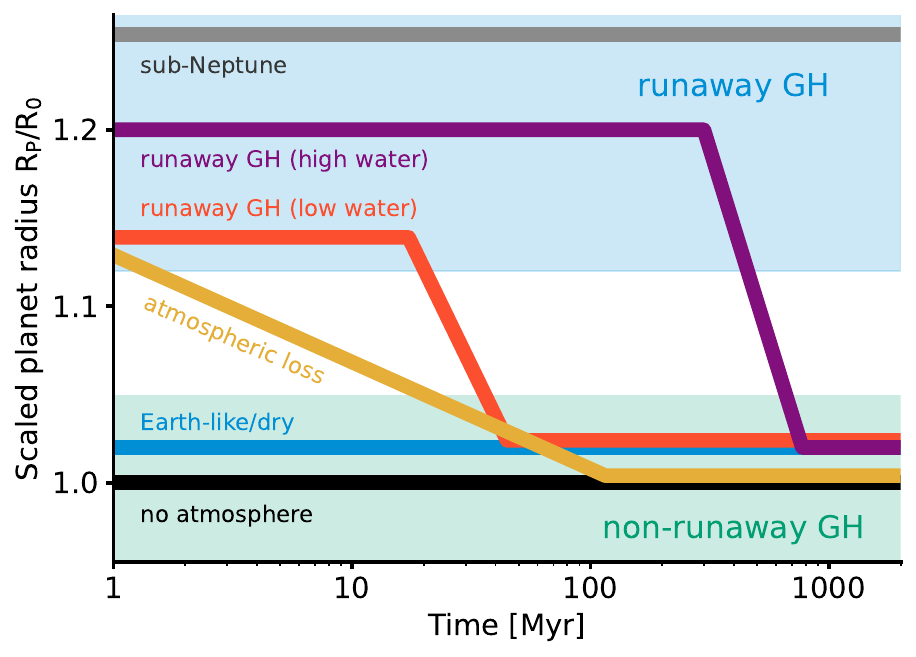}
        \caption{
            \textit{Top}: mass–radius relationships for a bulk water mass fraction $x_{H_2O}= \var{wrr}$ and different planet states. Green: planets with a solid mantle and no steam atmosphere. Dashed: planets with steam atmospheres. Blue: planets with steam atmospheres and including the effect of water retention in the melt.
                Steam atmospheres cause a significant radius increase, which is slightly reduced when water retention in the melt is considered.
            \textit{Bottom}: Radius evolution of different planet types, illustrating degeneracies and potential for confusion among planet classes. Shown is a schematic time evolution of the transit radius normalized to the atmosphere-free radius $R_\mathrm{0}$ for different scenarios. Planets can move between planet classes through processes such as atmospheric loss and desiccation, which ultimately ends a runaway greenhouse phase on a timescale dependent on a planet's water content.}
        \label{fig:radiusevolution}
    \end{centering}
\end{figure}

While we search for the signature of runaway greenhouse climates in demographic quantities such as average planet radii, the injected changes happen on the planetary level: We changed each planet's transit radius based on its individual set of properties and the associated predictions from steam atmosphere and water retention models.
Relevant properties are a planet's mass $M$, its net instellation $S$, and its bulk water inventory expressed as \rev{the total planetary} water mass fraction $x_{H_2O}$.
We consider the following cases (see Figure~\ref{fig:radiusevolution}):

\textit{Non-runaway} planets retain the radius assigned based on exoplanet occurrence rates (see Section~\ref{sec:syn_planets}).
This case serves as our null hypothesis.

\rev{We consider as \textit{runaway} planets those planets that absorb a stellar flux higher than a} dayside-averaged \rev{threshold instellation $S_\mathrm{thresh}$}.
\rev{For all planets absorbing an instellation exceeding} $S_\mathrm{thresh} = \var{injectedSthresh}$, we assume an inflated transit radius due to a steam atmosphere.
While the actual instellation threshold for a runaway climate depends on planetary albedo, surface gravity, and clouds\reviii{~\citep{2019Icar..317..583P,Turbet2021,Pierrehumbert2022,Turbet2023}}, this value was found to be a typical limit for the flux a planet can emit in a runaway greenhouse situation~\citep{Goldblatt2013,Kopparapu2013,Leconte2013,Hamano2015,Salvador2017,2019ApJ...875...31K,Boukrouche2021,2021JGRE..12606711L}.
\rev{This choice translates to about 1.18 times the instellation of present-day Earth with a fixed albedo of 0.3, which compares favorably to previous climate simulations~\citep{Leconte2013,Wolf2015}.
We adopt the same threshold instellation for all host star spectral types.}

To quantify the radius change, we applied the mass–radius relationships derived by \citet{Turbet2020} using a 1D~inverse radiative–convective model~\citep{Turbet2019}.
Their calculations rely on the same mass–radius relations for rocky interiors that we apply for our non-runaway planets~\citep{Zeng2016}.
For each planet above the instellation threshold, we assigned the predicted radius for the given water mass fraction and planet mass.

Nominally, the above models assume a \textit{dry melt} without dissolved volatiles.
Here, however, we consider a \textit{wet melt} magma ocean and take into account a radius decrease from \reviii{the } retention of water in the melt\reviii{ using the results published in \citet{Dorn2021}.}
The impact of the water distribution between melt and atmosphere on the change of the transit radius depends on the planet's mass and water content and is generally small compared to the radius inflation from the steam atmosphere.
We \reviii{obtained the} radius deviations between a wet magma ocean and a solid mantle \reviii{using the numerical values from} \citet{Dorn2021} \reviii{that assume} a tropopause pressure~$P_\mathrm{iso}=\SI{0.1}{\bar}$.
We then added the (in almost all cases negative) radius deviations to the inflated planet radii computed for the dry melt case.

We illustrate the mass–radius relations of the three cases in Figure~\ref{fig:radiusevolution} where a fiducial bulk water mass fraction of $x_{H_2O}= \var{wrr}$ is assumed. 
In the following, we only distinguish between the non-runaway greenhouse and wet melt scenarios.

To account for planets unable to sustain a steam atmosphere over extended time spans, as well as evolutionary effects such as desiccation through water photodissociation and H~escape (see \rev{bottom} panel of Figure~\ref{fig:radiusevolution}), we introduce a dilution parameter $f_\mathrm{rgh}$.
It represents the fraction of planets above the instellation threshold whose atmospheres are currently inflated due to a runaway greenhouse climate.
Our simulation setup is such that all planets receiving a net instellation $S < S_\mathrm{thresh}$ follow the non-runaway greenhouse relation, and a fraction $f_\mathrm{rgh}$ of the planets with $S > S_\mathrm{thresh}$ follow the wet melt relation.
The choice $f_\mathrm{rgh} < 1$ \rev{reflects the fact that planets -- despite a high irradiation --  can evade a runaway greenhouse climate.}
\rev{This situation may, for instance, arise in} the absence of an atmosphere or of volatiles that could form a steam atmosphere.
In the following, we test if and under what conditions this parametrization causes a demographic trend that is large enough to be detected with high significance.

\section{Exoplanet Survey Simulations and Hypothesis Testing}\label{sec:met_surveys-hypotests}
The survey module of \bioverse\ converts the synthetic planet sample into a set of uncertainty-laden measurements on a subset of that sample.
This task includes selection of the targets, application of detection biases, and conducting simulated measurements, all of which are specific to the particular survey.
For each planet-level measurement such as transit radius or instellation, we draw the measured value from a normal distribution centered on the true value with a standard deviation set by the survey's precision.
We then follow a Bayesian hypothesis testing approach to assess various realizations of simulated surveys in terms of their ability to detect and characterize the runaway greenhouse transition.

\subsection{Detection bias, target selection, and sensitivity}\label{sec:sensitivity} 
Not all transiting planets are detectable with the same likelihood and detection biases have an impact on the demographic measurements we are interested in.
A detailed characterization of the detection biases of individual missions would not be justifiable given the uncertainties of the theoretical predictions.
Instead, we derived generic observing limits that reflect the limitations of state-of-the-art transit surveys.

A successful transit detection requires a sufficient signal-to-noise ratio, which is sensitive to the achieved photometric precision. 
PLAnetary Transits and Oscillation of stars (\plato) is an ESA mission designed to characterize terrestrial planets in the habitable zones of Sun-like stars via long-term high-precision photometric monitoring of a sample of bright stars~\citep{Rauer2016}.
In line with this requirement, \plato\ is designed to enable the detection of a \SI{80}{\ppm} transit signal\rev{~\citep{Matuszewski2023,plato2017}}.
To reflect its sensitivity, we chose a minimum transit depth of \SI{80}{\ppm} as a detection limit and consider only measurements of planets exceeding this threshold.
We further exclude target stars with \gaia\ magnitudes $M_\mathrm{G} > \minMagnitude$.

The runaway greenhouse effect becomes obsolete both for very small instellations and where no atmosphere can be maintained due to proximity to the host star and resulting atmospheric erosion.
Ensuring to stay well clear of such regions, we clear our sample from all planets with a net instellation $S < \SI{10}{\watt\per\square\meter}$ or $S > \SI{2000}{\watt\per\square\meter}$.
We further consider only rocky planets with masses below \SI{2}{\Mearth}. 

\subsection{Measurements and their uncertainties}
Under real-world conditions, the planetary properties in question can only be probed with a finite precision that is specific to each exoplanet mission.
\plato's definition study report~\citep{plato2017} states precision requirements for planet radii~(\SI{3}{\percent}), planet masses through radial velocity~(RV) follow-up~(\SI{10}{\percent}), and stellar masses, radii, and ages~(\SI{10}{\percent}).
We adopted these estimates and assumed a \SI{10}{\percent} error on instellation measurements.

Since planetary bulk density $\rho \propto R_\mathrm{P}^{-3}$, we expect a stronger runaway greenhouse signal when measured through bulk density instead of transit radius.
We thus simulated measurements of planetary densities assuming the mass–radius relation defined above.
For uncertainties in bulk density measurements, we propagated the errors of the mass measurements assuming $\sigma_\mathrm{M_\mathrm{P}} = \SI{10}{\percent}$.

\subsection{Hypothesis tests}
We now turn to quantifying the ability of the simulated surveys to detect the habitable zone inner edge discontinuity and to constrain parameters associated to the runaway greenhouse transition.
To do this, we rely on a Bayesian hypothesis testing approach where we quantify the evidence of a hypothesis over another based on the (simulated) data.
For our specific problem, this implies comparing evidences for a demographic imprint of the runaway greenhouse effect to its absence.
    As a null hypothesis, we consider the case where the planetary radius distribution is independent of the instellation,
    \begin{equation}
        H_0(\theta, S) = \theta,
    \end{equation}
    where $\theta$ is the set of parameters defining the radius distribution.
    We further define an alternative hypothesis that describes radius changes due to runaway greenhouse climates and inflated steam atmospheres.
    As motivated above, this hypothesis takes the form of a step function in net instellation $S$, where the step occurs at the outer edge of the runaway greenhouse region.
    Our main observable shall be the average transit radius in the planet population on either side of this threshold.
    The runaway greenhouse hypothesis is then defined as
\begin{equation}\label{eq:rgh_hypo}
    H_{\mathrm{rgh}}(\theta, S) =
        \begin{cases}
            H_0, &  S \leq S_\mathrm{thresh}\\
            \langle R_\mathrm{P}\rangle (f_\mathrm{rgh},\Delta R_\mathrm{stm}, \Delta R_\mathrm{wtr}), &  S > S_\mathrm{thresh}.
        \end{cases}
\end{equation}
    Here, $f_\mathrm{rgh}$ is the fraction of planets above the instellation threshold experiencing a runaway greenhouse effect.
    $\Delta R_\mathrm{stm}$ and $\Delta R_\mathrm{wtr}$ are predicted radius changes from the steam atmosphere and water retention models, respectively.
    They are assumed to act additively on the planet radii and thus on their average $\langle R_\mathrm{P}\rangle $.

The only free parameter of the null hypothesis, which assumes the average transit radius to be independent of instellation, is the predicted mean radius $\langle R_\mathrm{P}\rangle $.
The functional form of the runaway greenhouse hypothesis is more complex: Besides the mean radius of planets outside the threshold $\langle R_\mathrm{P}\rangle_\mathrm{out}$, which is a nuisance parameter necessary to define the hypothesis, it relies on the threshold instellation for the ``step'' $S_\mathrm{thresh}$, the planetary bulk water mass fraction~$x_{H_2O}$, and the dilution factor $f_\mathrm{rgh}$.
For hypothesis tests based on bulk density instead of radius, we proceeded in the same way and substituted $R_\mathrm{P}$ by the bulk density $\rho$.

A sensible choice of priors is central for evidence estimation via nested sampling.
As the parameters of interest are poorly constrained by previous data, we used relatively uninformative priors to sample the entire physically plausible parameter space.
For $S_\mathrm{thresh}$, we chose a uniform prior in $[\prSmin,\prSmax]\, \SI{}{\watt\per\meter\squared}$.
We sampled $x_{H_2O}$ from a log-uniform distribution to imply scale-invariant ignorance.
Its boundaries $[\prWRRmin, \prWRRmax]$ are motivated by the water mass fractions covered by the geophysical models~(Section~\ref{sec:met_rghmodel}).
For $f_\mathrm{rgh}$, we chose a uniform prior in $[0, 1]$.
Finally, we adopted a broad, uniform prior for $\langle R_\mathrm{P}\rangle_\mathrm{out}$ bound by $[\prRmin, \prRmax] \, \mathrm{R_\oplus}$.
In the case of measuring bulk densities instead of transit radii, we drew uniformly from $[1, 6] \, \SI{}{\gram\per\centi\meter\cubed}$.

The measured radii $R_\mathrm{P, i}$ or bulk densities $\rho_\mathrm{i}$ cannot be directly used for the hypothesis tests as they include intrinsic scatter that is not caused by measurement errors.
$H_{\mathrm{rgh}}$ and $H_0$ should thus be tested against a statistical estimator that represents the population mean.
To avoid binning and the artificial patterns it may introduce, we chose to test our hypotheses against a simple moving average $SMA$ along the instellation axis with a window of size \windowsize\ centered around each measurement.\footnote{See Appendix~\ref{app:binnedstats} for a robustness test using a different estimator.}
We further computed the uncertainty of this moving average by propagating the individual measurement errors and applying a rolling standard error of the mean. 

As our procedure involves random sampling of the model parameters $\theta$, we need to define the probability of obtaining a data set given the model parameters, i.e., a likelihood function $\mathcal{L}$.
We assumed here that the individual moving averages $SMA_i$ are measured with a normally distributed uncertainty $\sigma_{SMA_i}$ and adopted a normal distribution
\begin{eqnarray}
    \mathcal{L}(SMA \mid \boldsymbol{\theta})= & \prod_{i}^{N} \frac{1}{\sqrt{2 \pi \sigma_{SMA_i}^{2}}} \\
    & \times \exp \left(-\frac{\left(SMA_i - H\left(\boldsymbol{\theta}, S_i\right)\right)^{2}}{2 \sigma_{SMA_i}^{2}}\right).
\end{eqnarray}
Here, $H\left(\boldsymbol{\theta}, S_i\right)$ corresponds to the functional form of the runaway greenhouse or null hypothesis.

\subsection{Bayesian model comparison}
We can now assess the relative plausibility of $H_{\mathrm{rgh}}$ and $H_0$ given the synthetic data we have generated, assigning equal a~priori~ probabilities to these models.
This is done by comparing the Bayesian evidence $\mathcal{Z}$ of the models, which we estimated with the nested sampling~\citep{Skilling2004} algorithm \emph{dynesty}~\citep{Speagle2020}.
We initialized the sampler with the priors defined above to let it estimate the evidence and sample the posterior distributions.
Our criterion to reject the null hypothesis is
\begin{equation}
\Delta \ln \mathcal{Z}  = \ln \mathcal{Z}_\mathrm{rgh} - \ln \mathcal{Z}_0  > 3.
\end{equation}

\section{Results}\label{sec:results}
\subsection{Statistical signature of the runaway greenhouse threshold}\label{sec:res_signature}
To characterize the population-level imprint of individual radius changes, we generated a generic planet population with an injected runaway greenhouse effect assuming a water fraction $x_\mathrm{H_2O} = \var{wrr}$.
\begin{figure*}
    \script{HnullHmo.py}
    \begin{centering}
        \includegraphics[width=\hsize]{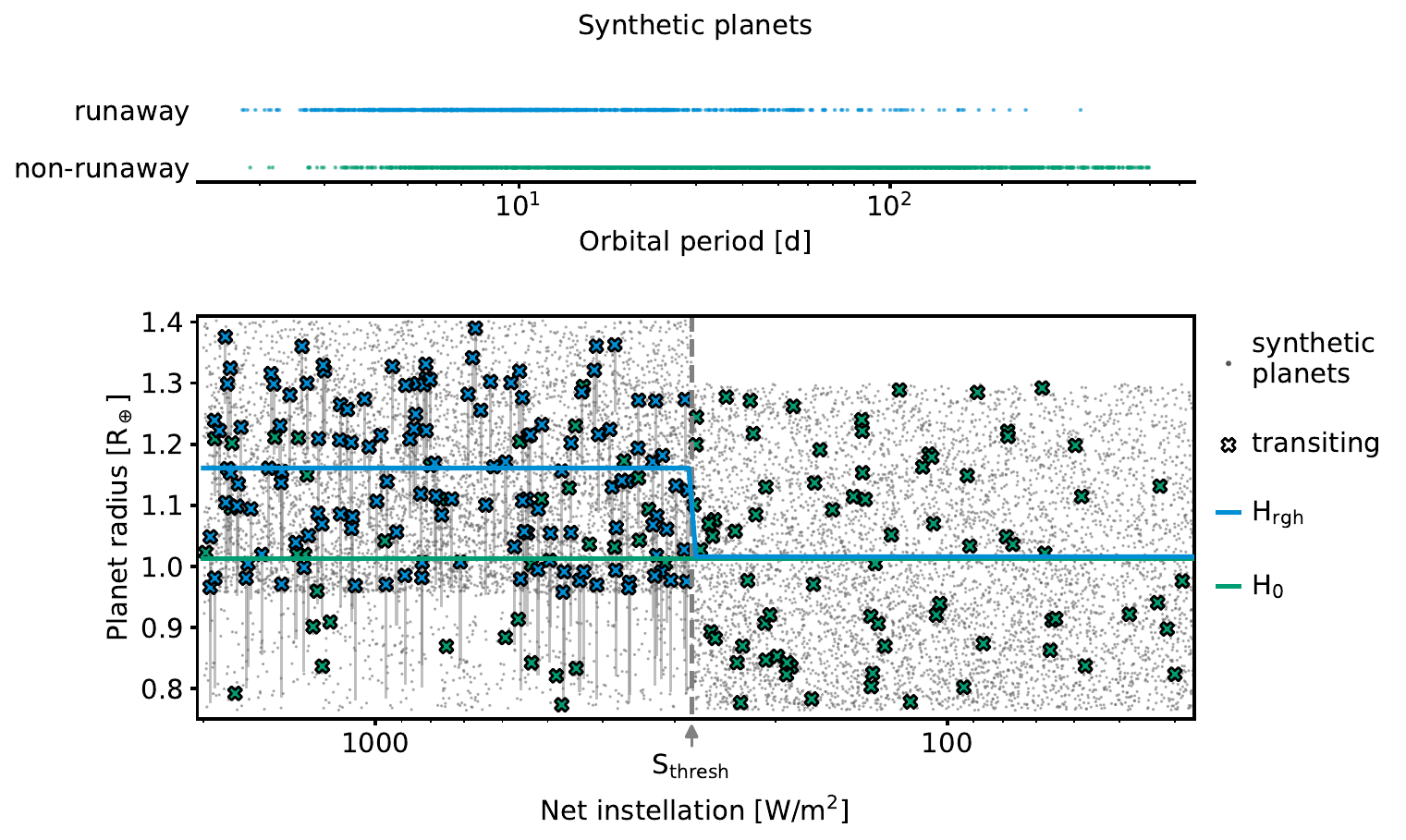}
        \caption{Synthetic planets above and below the runaway greenhouse threshold.
        \textit{Top}: Planet state as a function of orbital period. Planets with and without a runaway greenhouse climate mix and are not distinguishable in orbital period space.
        \textit{Bottom}: Transit radii of synthetic planets with injected radius deviation as a function of net instellation. Only the planets marked as transiting are observable.
            Above the runaway greenhouse threshold $S_\mathrm{thresh} = \var{injectedSthresh}$, some planets maintain their original radii (green crosses) whereas some have their transit radius inflated (blue crosses) by the amount indicated with gray lines.
        The sharp boundary at $S_\mathrm{thresh}$ causes a discontinuity in the average planet radius (blue line).
            This runaway greenhouse hypothesis can be tested against the null hypothesis $H_0$ (green line), where average radii are independent of instellation.}
        \label{fig:HnullHmo}
    \end{centering}
\end{figure*}
Figure~\ref{fig:HnullHmo} shows the resulting planetary radii.
When ordered in orbital period space, the different planet types overlap, diluting the demographic imprint.
With net instellation as an independent variable, planets above and below the runaway greenhouse threshold separate: The runaway greenhouse-induced radius inflation introduces a discontinuity of average planet radii and bulk densities as a function of stellar irradiation.
We also show the predictions of observable average planet radii from the statistical hypotheses defined above.
Within the runaway greenhouse regime, an average radius change of \avgRadiusChange\ occurs.
This pattern is consistent with the injected radius inflation as predicted from the atmospheric models (see Appendix~\ref{app:model-pop_comparison} for an investigation of the interplay between model predictions and our synthetic planet population).

\subsection{Testability of the runaway greenhouse hypothesis}\label{sec:res_testability}
\begin{figure}[ht!]
    \script{optimistic_RS.py}
    \begin{centering}
        \includegraphics[width=\linewidth]{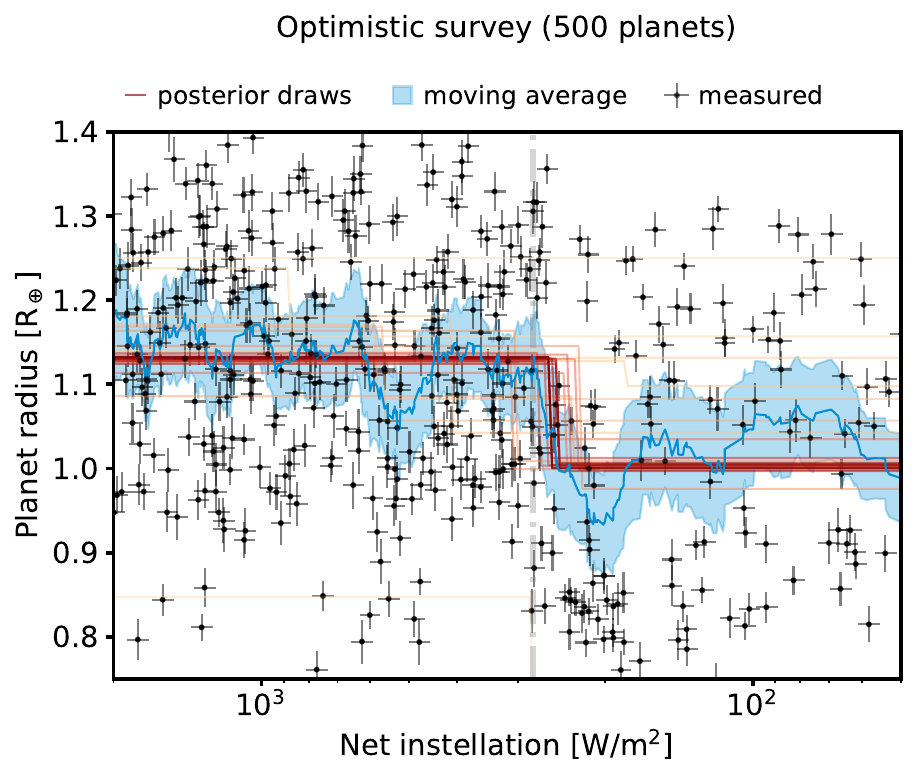}
        \caption{
        Detection of the runaway greenhouse threshold.
        From simulated radius and instellation measurements of a large ($N = \var{N_optimistic}$) survey, we compute the moving average (blue confidence intervals) and fit the runaway greenhouse hypothesis to it (Equation~\ref{eq:rgh_hypo}, random draws from the posterior in red).
            The pattern is detected with high significance.
        }
        \label{fig:optimistic_R-S}
    \end{centering}
\end{figure}
Figure~\ref{fig:optimistic_R-S} shows a prototypical statistical detection of the habitable zone inner edge discontinuity.
Before we study limiting cases of such a detection below, we first demonstrate the interpretation process based on an optimistic scenario where the sample of characterized planets is large ($N = \var{N_optimistic}$) and the measurement uncertainties are small ($\sigma_{R} = \sigmaR, \sigma_{S} = \sigmaS$).
Here, we assumed that the fraction of those planets irradiated stronger than $S_\mathrm{thresh}$ that have runaway greenhouse climates is $f_\mathrm{rgh} = \frgh$, and we chose a water mass fraction of $x_{H_2O} = \var{wrr}$ for each planet.
In this case, the habitable zone inner edge discontinuity was detected with high significance ($\Delta \ln \mathcal{Z} \approx 100$).

\begin{figure}[ht!]
    \script{cornerplot.py}
    \begin{centering}
        \includegraphics[width=\linewidth]{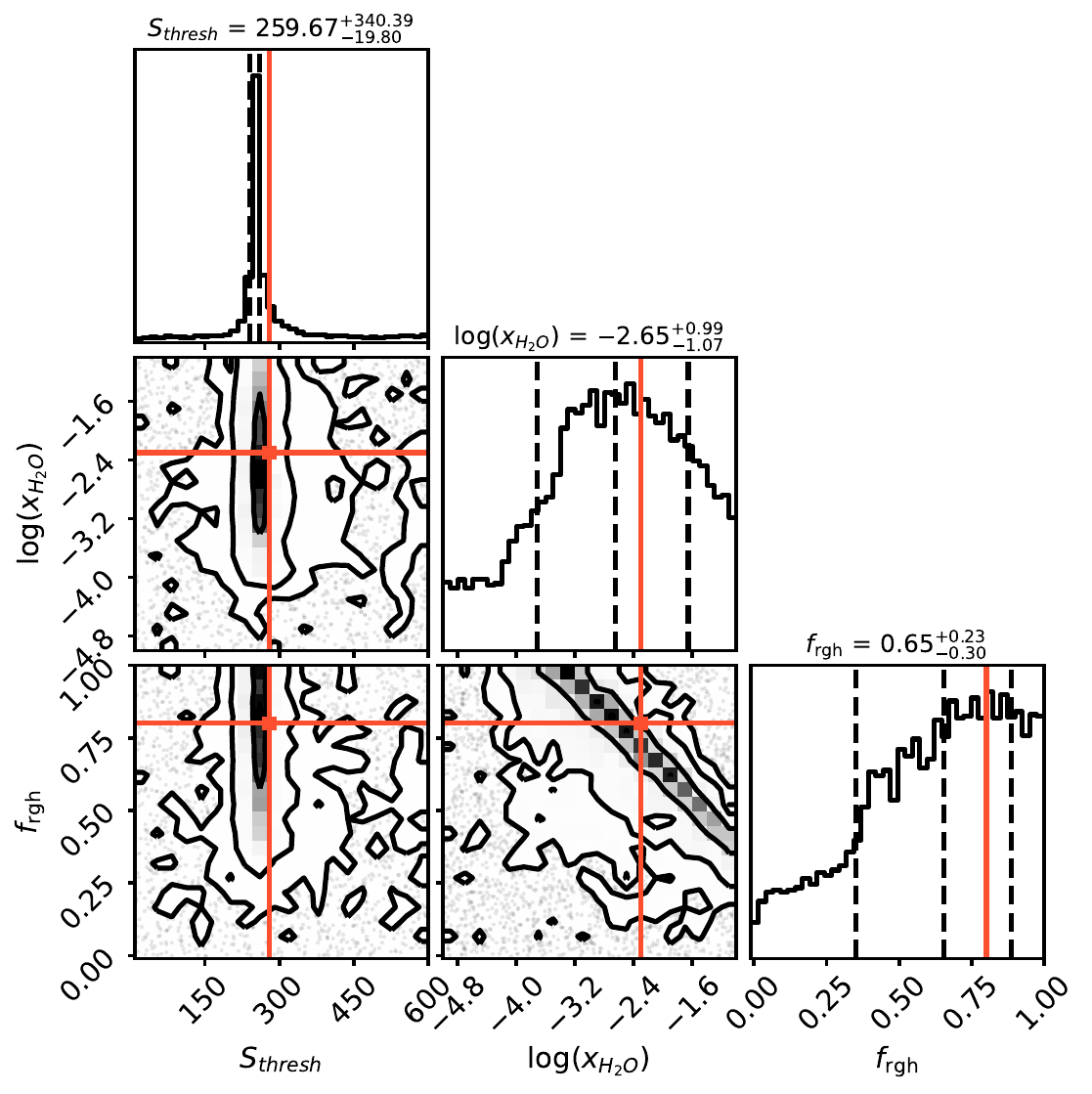}
        \caption{
            Retrieved posterior distribution of key parameters in the optimistic scenario. The density maps in each panel show relationships between and marginalized distributions of the threshold instellation $S_\mathrm{thresh}$, the \rev{bulk} water mass fraction $x_{H_2O}$, and the dilution factor $f_\mathrm{rgh}$ as they could be retrieved with a high-precision transit survey and a sample of \var{N_optimistic} planets. True values of the parameters for the injected effect are shown in orange. The threshold instellation can be reasonably constrained; the predominant water fraction and the fraction of planets with runaway greenhouse climates are degenerate.
        }
        \label{fig:cornerplot}
    \end{centering}
\end{figure}
With such a strong signal, we can attempt an inference of the parameters defining the injected effect.
Figure~\ref{fig:cornerplot} shows the posterior distributions of $S_\mathrm{thresh}$, $x_{H_2O}$, and $f_\mathrm{rgh}$ as determined by the nested sampler.
The threshold instellation can be accurately constrained. 
Both a higher water mass fraction and a higher dilution factor lead to larger average radii, thus these parameters are strongly correlated.

\begin{figure}[ht!]
    \script{optimistic_statpwr_H2O-f.py}
    \begin{centering}
        \includegraphics[width=\linewidth]{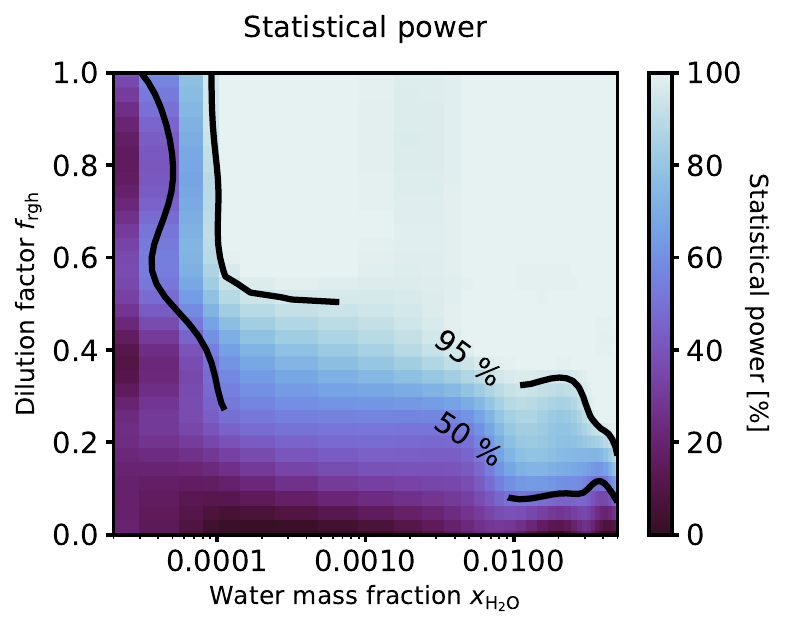}
        \caption{
            Statistical power of the runaway greenhouse hypothesis test as a function of model parameters.
            For a sample size $N = \var{N_optimistic}$, the color code shows the fraction of simulations resulting in a sound detection ($\Delta \ln \mathcal{Z} > 3$) for different combinations of \rev{bulk} water mass fraction and dilution factor.
            Higher values in either parameter result in a more reliable detection.
            For water mass fractions $\gtrsim 10^{-4}$, the statistical power largely depends on the fraction of greenhouse climate planets in the sample.
        }
        \label{fig:statpwr_H2O-f}
    \end{centering}
\end{figure}
Figure~\ref{fig:statpwr_H2O-f} explores the statistical power of the hypothesis test achieved in the above scenario for different combinations of the poorly constrained parameters $x_{H_2O}$ and $f_\mathrm{rgh}$.
It is highest for large water inventories and large dilution factors.
 For all but very low water fractions, $f_\mathrm{rgh}$ dominates this trend: It enters linearly into the average planet radius, whereas the contribution of $x_{H_2O}$ - as predicted by the geophysical models - is sublinear with a power-law exponent of $\sim 0.3$.
Within the framework of our model and as long as $f_\mathrm{rgh}$ is larger than $\sim 0.2$, a sample size of \var{N_optimistic} is sufficient for a $\SI{50}{\percent}$ detection rate even for water ratios as low as $10^{-3}$.

\subsection{Detecting the runaway greenhouse transition with \plato}\label{sec:res_samplesize}

    \begin{figure*}[ht!]
    \script{plato_grids.py}
        \begin{centering}
            \includegraphics[width=\linewidth]{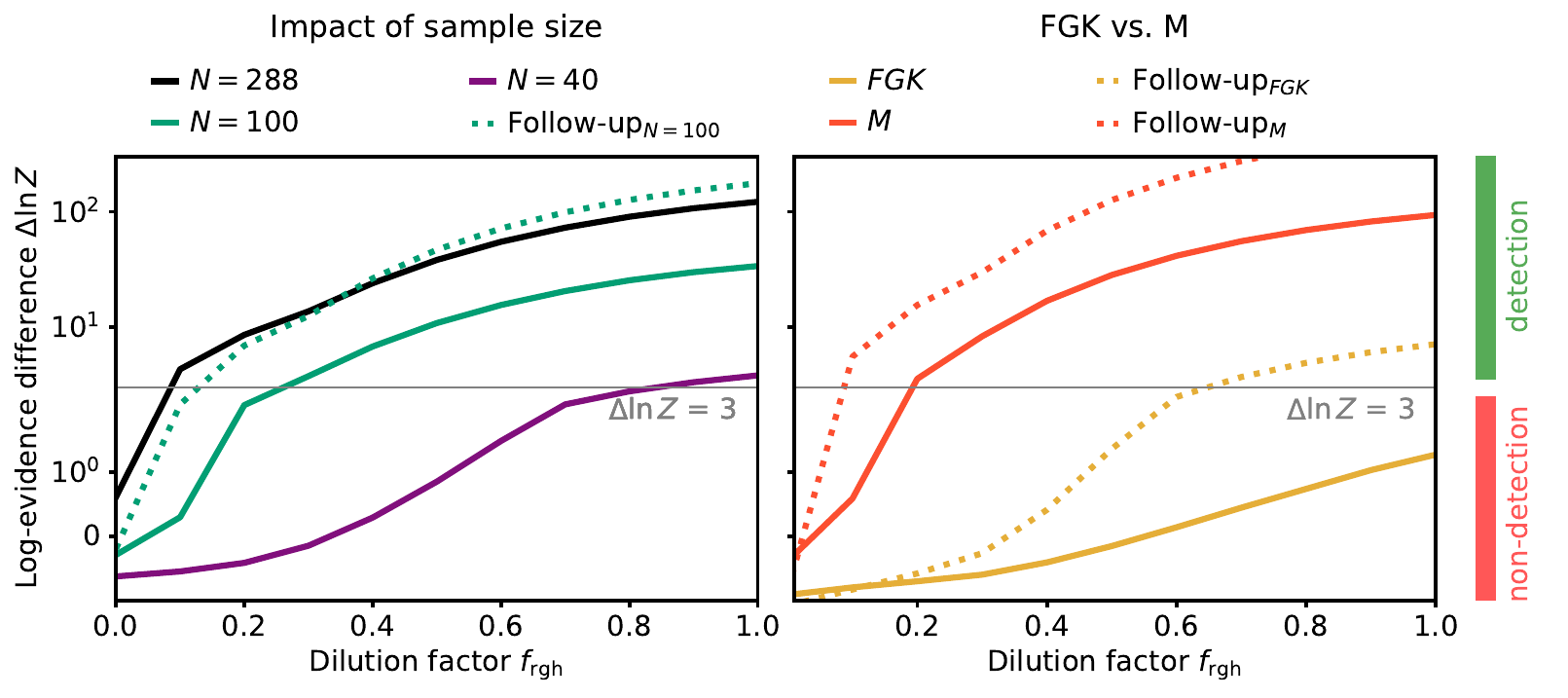}
            \caption{
            Expected delta-evidences as a function of the fraction of planets with runaway greenhouse climates for different versions of the \plato\ survey.
            The median values of randomized survey simulations are shown; $\Delta \ln Z > 3$ (gray horizontal line) is considered sufficient evidence to reject the null hypothesis.
            \textit{Left:} For a large planet yield of $N \approx \var{N_plato}$, even small dilution factors $\sim \var{platoMinfrgh}$ allow a detection.
                A sample of 100 planets is sufficient if their masses are constrained to within \SI{10}{\percent} (dotted green line).
                Without such follow-up measurements, sufficient diagnostic power can only be achieved with this sample if $f_\mathrm{rgh} \gtrsim \var{platoMinfrghHundred}$.
               Even smaller samples are unlikely to yield a significant detection.
            \textit{Right:} Evidences when only FGK or only M~dwarfs are considered.
                Only M~dwarfs host enough planets on both sides of the threshold instellation to allow a reliable detection of the habitable zone inner edge discontinuity.
            }
            \label{fig:plato_fgrid}
        \end{centering}
    \end{figure*}
To simulate the transit survey of ESA's \plato\ mission, we considered a volume-limited sample with a size according to projections and including all stellar spectral types.
Given \plato's expected radius precision, we find that a yield of $\sim$\var{N_plato} is needed for a significant detection if the fraction of runaway greenhouse planets $ f_\mathrm{rgh} = \var{platoMinfrgh}$ (see Figure~\ref{fig:plato_fgrid}).
The minimum needed fraction rises to \var{platoMinfrghHundred} for $N = 100$.
For much smaller samples, only an optimistically strong signal is likely to be detected.

\subsection{Statistical power of different mission designs}\label{sec:statpower_missions}

\subsubsection{Additional planet mass measurements}\label{sec:res_followup}
Comparing a measurement of the habitable zone inner edge discontinuity in radius space with a measurement in density space (which requires planetary mass measurements), a stronger detection occurs in the latter case:
With an optimistic choice of geophysical parameters (see Section~\ref{sec:res_testability}), the average measured radius change is \SI{+15}{\percent} whereas the average density change is \SI{-33}{\percent}.

Figure~\ref{fig:plato_fgrid} shows that investigations of the discontinuity are more constraining when radius measurements can be augmented with mass measurements: At unchanged sample size, the difference in Bayesian evidence can be up to an order of magnitude larger.
Consequently, we achieve a statistically significant detection with smaller samples or lower dilution factors $f_\mathrm{rgh}$.
A density-based survey of 100 targets is roughly equivalent to a radius-based survey of \var{N_plato} targets.
At $N=100$, pure radius measurements require $f_\mathrm{rgh} \gtrsim \var{platoMinfrghHundred}$ whereas bulk density measurements enable a detection from $f_\mathrm{rgh} \gtrsim \var{platoMinfrghHundred_rho}$.

\subsubsection{Dependence on host star spectral type}\label{sec:results_FGK_M}
Since the incident radiation at a given orbital distance depends on the spectral type of the host star, the relative number of planets on either side of $S_\mathrm{thresh}$ is different for FGK and M~dwarfs.
We tested the detectability of the habitable zone inner edge discontinuity when only FGK or only M~dwarfs are considered (see Figure~\ref{fig:plato_fgrid}).
\rev{All other parameters of the climate models are kept the same.}
The samples are volume and magnitude-limited to reflect the target counts of \plato's provisional Long-duration Observation Phase fields~\citep[$15996$ FGK stars in the P1 and P2 samples, $33948$ M~stars in the P4 sample, ][]{Nascimbeni2022}.
The resulting M~dwarf planet sample is significantly larger with $\var{N_M}\pm \var{N_M_err}$ planets compared to $\var{N_FGK}\pm \var{N_FGK_err}$ planets in the FGK sample.
No significant detection is possible in the pure FGK sample, independent of the assumed geophysical parameters.
In the M~dwarf sample, the evidence threshold is reached around $f_\mathrm{rgh} \sim 0.2$, similar to the case above where all spectral types are considered.

\subsubsection{Constraining the threshold instellation}\label{sec:res_constrain-S_thresh}
\begin{figure}[ht!]
    \script{plato_Sthresh_grid.py}
    \begin{centering}
        \includegraphics[width=\linewidth]{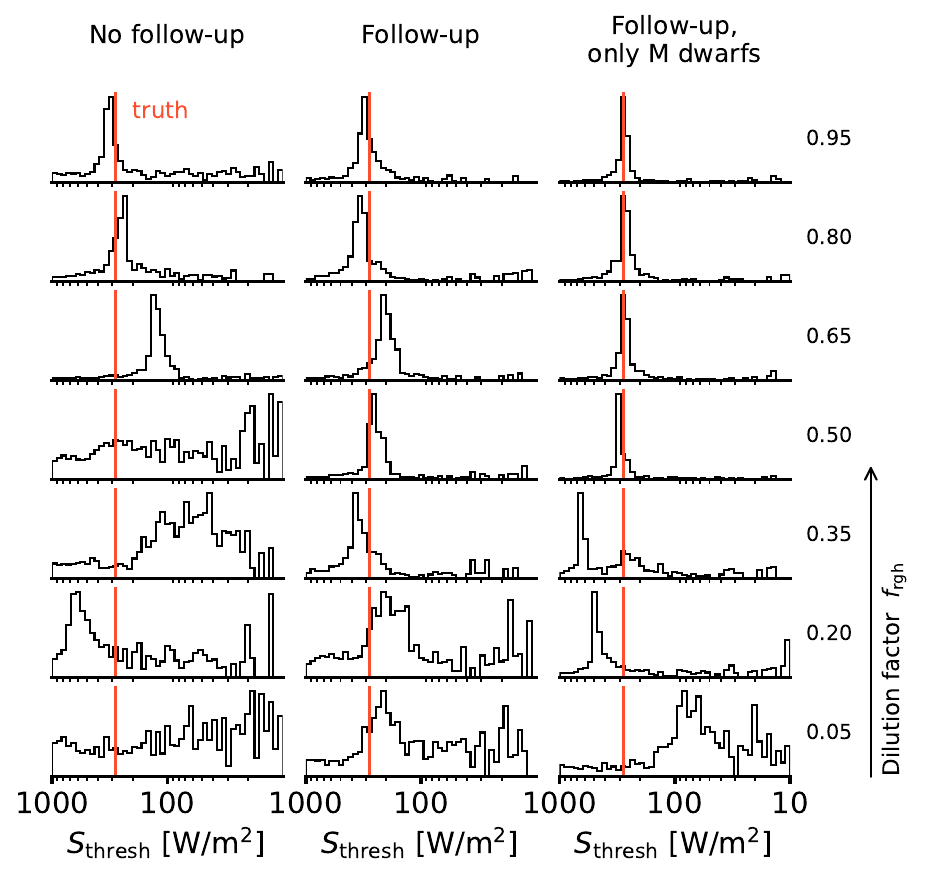}
        \caption{
            Retrieved posterior distributions of the threshold instellation for different survey realizations.
            All cases assume $x_{H_2O} = \var{wrr}$ and a planet sample size $N=100\pm10$; The fraction of planets with runaway greenhouse climates varies across rows.
            Orange lines show the true value of the injected signal.
            Accuracy and precision of the constraint on $S_\mathrm{thresh}$ generally improve with higher $f_\mathrm{rgh}$.
            Bulk density-based inferences improve the constraints, and M~dwarf samples yield the highest accuracy and precision.
        }
        \label{fig:posterior_surveys}
    \end{centering}
\end{figure}
\rev{Although the exact location of the runaway greenhouse transition depends on the stellar energy actually absorbed by the planet, }a measurement of $S_\mathrm{thresh}$ \rev{would be a} key constraint resulting from a detection \rev{of the habitable zone inner edge discontinuity}.
Here, we assess the ability of different mission concepts explored above to constrain this parameter.
Figure~\ref{fig:posterior_surveys} shows posterior distributions from a grid of inferences together with the true values of the injected signal for different fractions $f_\mathrm{rgh}$.
We consider three cases: only radius measurements, radius and mass measurements, and radius and mass measurements of only planets orbiting M~dwarfs.
The simulations otherwise represent the simulated \plato\ example described above.
We chose a planetary sample size of $N=100\pm10$, as this was found to be a threshold case in Section~\ref{sec:res_samplesize}.

We find that retrievals from radius measurements alone require high fractions of greenhouse climate-bearing planets to achieve an accurate constraint on $S_\mathrm{thresh}$:
dilution factors $f_\mathrm{rgh} \gtrsim 0.5$ yield posterior probability distributions that are condensed at the order of magnitude of the true value; accurate constraints to within $\pm 0.25\,\mathrm{dex}$ of the truth are reached only from $f_\mathrm{rgh} \approx 0.8$.

In contrast, if planet masses are available and the hypothesis test is conducted in bulk density space, useful constraints emerge already from about $f_\mathrm{rgh} \approx 0.35$.
Accuracy and precision of the retrievals are improved.

A sample containing only planets around M~dwarf yields still better performance and result in the overall best accuracy and precision.
The reason is that these planets contribute most to the statistical power of the retrieval.

\section{Discussion}\label{sec:discussion}

\subsection{Statistical imprint of exoplanet climates}
We showed in Section~\ref{sec:res_signature} that injecting the theoretically predicted radius inflation effect into a synthetic planet population following the currently known demographics leaves a distinct pattern in the radius and density distribution as a function of instellation.
The transit radius inflation was suggested before as an observational diagnostic to probe the runaway greenhouse transition~\citep{Turbet2019}, and our simulations show quantitatively how the contributions of the individual radius changes combine to create a significant demographic signal.
Its strength depends on largely unknown factors such as the planetary volatile content, but turned out to be well in the detectable range under reasonable assumptions of these factors.

Our finding that the expected habitable zone inner edge discontinuity is strong except for extremely low water mass fractions gives reason for optimism regarding its detection.
It also presents a potential for constraints on the water inventory of terrestrial planets in the case of a non-detection.
Such constraints will provide insight into whether the initial water content of rocky planets is varied by systematic inter-system effects~\citep[e.g.,][]{Raymond2004,Mulders2015b,Sato2016,Lichtenberg2019,2021ApJ...913L..20L,2022ApJ...938L...3L} or if there is a predominant pattern of volatile-enrichment across planetary systems that is only modified by intra-system effects such as planet migration~\citep{Schlecker2021} or atmospheric escape~\citep{Owen2016}.
Ultimately, such measurements will shed light on how delivery and loss effects during rocky planet formation and evolution shape the diversity of exoplanetary climates.
Overall, probing the runaway greenhouse discontinuity appears to be the most promising approach toward a first empirical test of the habitable zone concept.

\subsection{Detectability of the habitable zone inner edge discontinuity}
We showed in Section~\ref{sec:res_testability} that under favorable conditions, a sufficiently large (\var{N_optimistic} planets) photometric survey is likely to detect the demographic imprint of the runaway greenhouse transition and accurately constrain the associated threshold instellation.
Of course, even assuming capable instrumentation and an optimal survey design, our current models may not capture fully the complexity of trends in the geophysics and demographics of small planets.
The detectability of the runaway greenhouse threshold is thus a function of both survey design and the diverse outcomes of rocky planet formation and evolution.
In the following, we will explore the key factors that may influence the emerging demographic signal.

\subsubsection{Key factors influencing tests of the runaway greenhouse hypothesis}\label{sec:dis_keyfactors}
What influences the probability of correctly rejecting a false null hypothesis?
We \rev{identified six drivers} of the diagnostic power for detecting the runaway greenhouse transition with a transit survey.
\rev{Many of these factors directly influence the fraction of planets currently in runaway greenhouse states, which in our model is represented by the dilution factor $f_\mathrm{rgh}$.}
\begin{itemize}
    \item Occurrence rate of planets forming steam atmospheres
    \item Planetary evolution and duration of the steam atmosphere phase
    \item Prevalent water inventory
    \item Size and composition of the planetary sample
    \item Radius measurement precision
    \item Availability and precision of mass measurements
\end{itemize}
We will now briefly explore the above drivers.

\textit{Occurrence rate of planets forming steam atmospheres}: The runaway greenhouse climate relies on sufficient amounts of atmospheric water vapor that can act as a greenhouse gas.
\revii{It was shown that} already about $\SIrange{\sim 10}{20} \, \mathrm{bar}$ of water vapor -- corresponding to a minor fraction of one Earth ocean and thus the lower limit of water on Earth -- is enough to sustain sufficiently high surface temperatures to keep the planet in a magma ocean stage~\citep{Boukrouche2021,Lichtenberg2021c}.
\revii{However, recent simulations accounting for radiative–convective profiles of near-surface atmospheric layers suggest the existence of cooler pure-steam runaway greenhouse states that do not necessarily yield a molten surface~\citep{Selsis2023}.
Climate state and atmospheric structure, as well as the fraction of planets fulfilling the requirements of a steam atmosphere} have an impact on the amplitude of the demographic imprint of the runaway greenhouse transition.
From a planet formation perspective, the incorporation of water into planets in the terrestrial planet zone is a standard expected outcome~\citep[e.g.,][]{2019PNAS..116.9723Z,Venturini2020,Emsenhuber2021b,Schlecker2021,Burn2021}.
But while commonly considered volatile delivery channels suggest a fraction of planets to be volatile-poor, the incorporation of hydrogen into even the driest planetary materials known in the solar system\rev{~\citep{McCubbin2019,2020Sci...369.1110P,2021PSJ.....2..244J}} suggests that hydrogen is present in all rocky planets upon formation.
Accreted hydrogen in nominally dry planetary materials reacts with mantle oxygen to form substantial amounts of water inside of the planet during the magma ocean phase~\citep{Ikoma2018,2021ApJ...909L..22K,2020MNRAS.496.3755K,2022NatAs...6.1296K}.
Enhanced equilibration between the core, mantle, and atmosphere during magma ocean evolution of rocky exoplanets further enhances this process~\citep{2021ApJ...914L...4L,Schlichting2022}.
Therefore, even nominally dry planets generate substantial amounts of water during formation and early evolution.
Reduced heating from short-lived radionuclides in extrasolar planetary systems increases the expected water abundance in exoplanet systems further, in particular for M~dwarf systems~\citep{2022ApJ...938L...3L}.

\textit{Planetary evolution and duration of the steam atmosphere phase}: An inflated steam atmosphere can only be sustained until the planet has lost its water.
Depending on host star spectral type, planetary mass, and composition, planets can spend from a few Myr to several Gyr in runaway greenhouse climates~\citep[][]{Hamano2015,Luger2015}, and only planets observed during this phase will contribute to the habitable zone inner edge discontinuity.
However, the delivery uncertainty is much greater than the predicted loss rates of water by atmospheric escape, which typically is limited to in total a few tens of terrestrial oceans per Gyr~\citep{2018AJ....155..195W} on even the most irradiated exoplanets~\citep[see discussion in][]{2022ApJ...938L...3L}.
\rev{The finite duration of steam atmosphere phases is a main factor for diluting the habitable zone inner edge discontinuity, and it leads to an increased likelihood of detecting this imprint for younger planetary systems.
Preferential selection of younger systems could therefore be of advantage in a targeted survey.}

\textit{Prevalent water inventory}: The magnitude of radius change at the runaway greenhouse threshold is sensitive to the water mass fraction.
As a result, the statistical abundance of water in terrestrial planets impacts the strength of the demographic pattern:
The higher the water content and the higher the fraction of planets in runaway greenhouse climates, the greater the likelihood that a discontinuity is detectable.
However, dissolution into the magma ocean and atmospheric inflation show non-linear coupling\rev{:
the more water there is in the atmosphere, the more will be dissolved~\citep{Dorn2021}, leading to a convergence in radii until saturation.
Most important for the present work, however, is the qualitative dichotomy between sub-runaway and runaway planets, which outcompetes radius variations within each of these climate states due to a changing water mass fraction~\citep{Turbet2020}.
If a planet hosts a significant water inventory that is not easily stripped by atmospheric escape~\citep{Johnstone2020}, then the statistical power of our model is high (see Figure~\ref{fig:statpwr_H2O-f}).
}

\textit{Size and composition of the planetary sample}: The significance of a statistical trend increases with a larger sample size.
In addition, the sample must include planets on both sides of the instellation threshold.
This is only likely for low-mass host stars due to the strongly distance-dependent detection bias associated with the transit method and the temporal coverage of upcoming transit missions.

\textit{Radius measurement precision}: The more precise individual planet radii can be determined, the more pronounced the discontinuity will be.
Good \textit{accuracy} is less important, as long as it does not have a systematic error scaling with stellar irradiance.

\textit{Availability and precision of mass measurements}: For simple geometric reasons ($\rho \propto R^{-3}$), the expected discontinuity at the habitable zone inner edge is stronger when measured in bulk density than it is in planet radius space.
If transiting planets can be followed up to obtain mass measurements, the statistical significance increases.

Besides these main factors, uncertainties in the measured instellations can influence the result, although they are typically small due to the very precise orbital period measurements available for transiting planets.
This can be different for young host stars when their ages cannot be well constrained; in particular, the long pre-main sequence phase of M~dwarfs shows a large variation in bolometric luminosity~(see Figure~\ref{fig:luminosity_tracks}).

\subsubsection{False positive scenarios}\label{sec:dis_falsepositive}
Runaway greenhouse climates are not the only physical mechanism that may cause a change in transit radius for a subset of planets.
Alternatives include atmospheric loss due to either photoevaporation through high-energy radiation by the host star~\citep[e.g.,][]{2012ApJ...753...66I,Owen2013,Jin2014,Mordasini2020a} or due to residual heat from the planet's interior shortly after formation~\citep{Ginzburg2016b,Ginzburg2018,Gupta2019}.
Both processes are being discussed as potentially sculpting the observed bimodality in the radius distribution of small exoplanets~\citep{Fulton2017,VanEylen2018}, and both lead to a decrease of planet radius for planets close to their host star~\citep{Pascucci2019,Bergsten2022}.
This is distinct from the radius inflation introduced by runaway greenhouse climates.
For example, the innermost planet in the \mbox{K2-3} system has an increased radius compared to its outer siblings, contrary to what would be expected from atmospheric escape~\citep{2022AJ....164..172D}.

Other false positive contributions may stem from potential unknown occurrence rate gradients in radius-instellation space, especially if these variations are similar to the expected habitable zone inner edge discontinuity.
Although an abrupt pattern at the expected location of the transition seems unlikely, examples of steep occurrence rate density changes exist.
An example is the ``Neptune desert'', a triangular region in period-radius space of low planet occurrence~\citep{Szabo2011,Mazeh2016,Dreizler2020b}.
The shape of this region is such that smaller planets become less frequent the closer to the star they are, which to some degree resembles the pattern introduced by the instellation dependency of the runaway greenhouse transition.
However, the Neptune desert occurs at smaller orbital periods and is sensitive to the planet radius~\citep{Szabo2011}, which is not expected for the runaway greenhouse transition.

\citet{Luque2022} found that small planets orbiting red dwarfs can be classified into three density regimes with a particularly strong separation between planets consistent with a pure rocky and those consistent with a water-rich composition.
This trend does not represent a false positive scenario for the habitable zone inner edge discontinuity, since no strong dependency on instellation has been found or is expected.
A population of ``water worlds'' with low bulk densities on a wide range of orbits would merely attenuate the statistical runaway greenhouse imprint.
If the dichotomy forms primordially through migrated planets that accreted from different regions of their protoplanetary disk~\citep{Venturini2020,Burn2021,Schlecker2021,Schlecker2021b} or from inter-system variations in the desiccation of volatile-rich planetesimals~\citep{Lichtenberg2019,2021Sci...371..365L,2021ApJ...913L..20L,2022ApJ...938L...3L,2023NatAs...7...39B}, systems of all ages can be affected by this attenuation.

\subsubsection{Atmospheric spectral signatures}\label{sec:dis_atmospheres}

We currently see four potential lines of discriminating atmospheric signatures associated with runaway climates in the exoplanet population.
(i) Detecting the atmospheric windows of water vapor in the near- to mid-infrared.
In a runaway greenhouse atmosphere, absorption is dominated by the opacity of water vapor, which has two prominent spectral features: one at 3.5--4.5 $\mu$m, and one between 8--20 $\mu$m~\citep[e.g.,][]{Boukrouche2021}.
Probing these features requires an instrument covering these wavelength ranges, for instance ELT~METIS~\citep{Brandl2021}, JWST~MIRI~\citep{Rieke2015}, or future missions such as the Large Interferometer For Exoplanets~\citep[\life,][]{2019A&A...621A.125B,Quanz2022,2022A&A...664A..22D}.
Potentially the two-band filter capabilities of \plato\ may offer insight into particularly pronounced spectral features in this runaway greenhouse regime~\citep{2020SSRv..216...98G}.
\rev{However, detecting water features in steam atmospheres may be obscured by high-altitude clouds that could potentially mute a transit signal as measured from peak to trough of an observed spectral line\reviii{~\citep{Suissa2020,Fauchez2019,Turbet2023}}.
A better understanding of the role of clouds in this problem may be critical for observational constraints on the runaway climate of individual exoplanets.}
(ii) Post-runaway planets may be detectable through O$_{3}$ absorption due to build-up of abiotic oxygen, leftover from photochemical dissociation of water, and hydrogen loss~\citep{2014ApJ...785L..20W,Luger2015}.
(iii) Water loss in runaway greenhouse episodes would increase the atmosphere's D/H isotopic ratio, akin to the enhancement in Venus' present-day atmosphere~\citep{2019JGRE..124.2015K,2021JGRE..12606643K}.
This may be detectable in high-resolution observations focusing on isotope-sensitive transitions~\citep{2019AJ....158...26L,2019A&A...622A.139M}.
(iv) Disequilibrium chemistry in tidally-locked runaway planets.
It has been suggested that the atmospheric depth and the presence or absence of oceans on ``sub-Neptunes'' could be probed via the abundance of species that are photochemically destroyed in the upper atmosphere, and replenished from either thermochemical layers~\citep{2021ApJ...914...38Y,2021ApJ...922L..27T} or at an ocean-atmosphere interface~\citep{2019ApJ...887..231L,2021ApJ...921L...8H}.
Atmospheric nitrogen and carbon compounds can be partitioned into magma~\citep{2022E&PSL.59817847G,2022PSJ.....3...93B}.
Therefore, it may be possible to discern the presence of an underlying magma ocean if the presence of an atmosphere on a rocky exoplanet can be confirmed.
This has been suggested to be done via (v) eclipse photometry~\citep{Mansfield2019,2019ApJ...886..140K} through the presence of a high albedo, which is expected to differ from the crystallized rock of a solidified magma ocean~\citep{2020ApJ...898..160E,Fortin2022}.
\rev{Spectral information from atmospheres of highly irradiated planets may also help to distinguish classical runaway greenhouse states from other climate regimes, such as a moist bistability~\citep{Leconte2013a}.}

\subsubsection{Detection in multi-planet systems}
As suggested by \citet{Turbet2019}, an alternative approach for detecting the runaway greenhouse-induced radius inflation is to search for its ``local'' imprints in multi-planet systems.
Systems harboring planets on both sides of the transition, such as \mbox{TRAPPIST-1}~\citep{Gillon2016a,Gillon2017a,Luger2017c,Agol2021}, \mbox{K2-3}~\citep{2022AJ....164..172D}, or \mbox{Kepler-138}~\citep{2022NatAs.tmp..269P}, may show the predicted abrupt radius and density change, provided the initial volatile content was sufficient and complete desiccation has not yet occurred.
While degeneracies remain in interpreting bulk density fluctuations within individual systems~\citep[e.g.,][]{Turbet2020,Dorn2021}, the detection of a consistent pattern in several such systems could be a convincing statistical evidence of the runaway greenhouse transition.
The current sample of suitable systems is sparse: The California-Kepler Survey catalog~\citep{Fulton2018} contains only six planets with instellations $< 2\, S_\oplus$ and smaller than \SI{2}{\rEarth} in five multi-planet systems.
Future additions to the multi-planet sample through missions such as \plato\ are needed.

\subsection{Diagnostic power of near-future exoplanet missions}\label{sec:dis_samplesize}
Confirming or disproving the predicted habitable zone inner edge discontinuity will depend on the significance with which the null hypothesis can be excluded, which is a function of instrumentation and survey strategy.
As discussed in Section~\ref{sec:dis_keyfactors}, key drivers from a mission design perspective are sample size, photometric precision, and the availability of planets around low-mass host stars.
We found that along these axes, \plato\ will be the most favorable among the upcoming transit missions.
The \plato\ team has released an estimate on the number of exoplanets that will be characterized in the course of the main survey mission. 
With an expected transit radius precision of \SI{3}{\percent}~\citep{plato2017} for hundreds of planets~\citep{Rauer2021}, the \plato\ mission is comparable to the optimistic survey (see Section~\ref{sec:res_testability}) in terms of sample size and precision.
If successful, it should readily detect the predicted statistical imprint or, in case of a non-detection, provide strong upper limits on the occurrence rate of runaway greenhouse planets.
The latter depends on the lifetimes of runaway greenhouse phases, which are a function of the initial water inventory of the planets~\citep{Hamano2015}.
Overall, it seems feasible to derive the typical water content of low-mass exoplanets from these occurrence estimates.

What other planned missions are suited to probe the habitable zone inner edge discontinuity?
\kepler\ and \ktwo\ have contributed a large number of discovered terrestrial-sized planets, but few of them are in the habitable zone and their host stars are typically too faint for RV follow-up with current instrumentation~\citep{Dressing2015}.

Similarly, the Transiting Exoplanet Survey Satellite~\citep[\tess,][]{Ricker2014a} planet sample lacks temperate, small planets around bright host stars~\citep{Ment2023}, as was expected from planet yield calculations~\citep{Barclay2018}.
As of March 22, 2023, the NASA Exoplanet Archive\footnote{\url{https://exoplanetarchive.ipac.caltech.edu}} lists 40 \tess\ candidate or confirmed planets smaller than \SI{4}{\rEarth} with lower estimated instellation than Earth's.

The ongoing \cheops\ mission was designed as a follow-up mission to search for transits of planets discovered with other techniques, in particular with radial velocity measurements~\citep{Benz2021}.
As such, it will provide precise radius constraints on a sample of small planets; however, only a small number of planets with orbital periods \SI{> 50}{\day} are being observed.
This largely limits \cheops' coverage to planets within the runaway greenhouse regime, preventing a detection of the transition.

As \cheops, the Atmospheric Remote sensing Infrared Exoplanet Large survey~\citep[\ariel,][]{Puig2016} will be a follow-up mission that is not designed to provide a large number of new radius measurements.
\ariel's primary targets are larger planets in the range of sub-Neptune to Jupiter-like planets.
We thus do not expect a significant contribution to statistically exploring the inner edge of the habitable zone for Earth-sized planets.

While not primarily designed to detect transiting planets, the Galactic Bulge Time Domain Survey of the \rst~\citep{Spergel2015} is expected to yield $\sim 10^5$ transiting planets on short orbits and constrain their radii in the course of its mission~\citep{Montet2017}.
$\mathcal{O} (1000)$ planets smaller than Neptune could be found around early to mid-M~dwarfs, however, only a small fraction of them will reach into the habitable zone~\citep{Tamburo2023}.
We thus conclude that the \rst\ could provide a useful sample to explore the runaway greenhouse transition, albeit with a predominant focus on water-rich (sub-)Neptunes~\citep[e.g.,][]{Pierrehumbert2022}.

Looking further ahead, the \nautilus\ Space Observatory concept~\citep{Apai2019} represents a statistical mission able to provide precise radius measurements of a large sample ($\sim \num{1000}$) of small exoplanets.
It employs a constellation of $\sim \num{35}$ large-diameter ($D \sim \SI{8.5}{\meter}$) telescopes using ultralight diffractive-refractive optical elements~\citep{Milster2020} with the primary goal to study the atmospheres of transiting exoplanets.
Operating in an array mode, \nautilus\ would achieve the equivalent light-collecting area of a \SI{50}{\meter} telescope.
Its expected \SI{1}{\ppm} photometric precision~\citep{Apai2022} would enable precise radius measurements of a large sample, also through a low number of required visits per object.
If realized, \nautilus\ will be a valuable instrument for characterizing the runaway greenhouse transition.

Other missions have been proposed that focus on characterizing exoplanet habitability, most notably the \hwo\ concept, which will build on the two precursor direct imaging concepts LUVOIR~\citep{LUVOIR2019} and HabEx~\citep{Gaudi2020c}.
Similar science objectives are pursued by the \life\ initiative~\citep[][]{Quanz2022}, a mission concept utilizing a space-based mid-infrared nulling interferometer.
Direct imaging surveys do not directly measure planetary radii and are primarily useful for providing context through atmospheric measurements of individual planets.
However, because mid-infrared retrievals feature reduced degeneracy between cloud albedo and changes in surface area, the planet radius can be constrained in mid-infrared wavelengths~\citep{2018ExA....46..543D,2021ExA...tmp..118Q}.
Mid-infrared direct imaging techniques, in particular, enable to study much deeper atmospheric layers than possible in reflected light~\citep{Wordsworth2022}.
Hence, the atmospheric structure can be retrieved for a wider variety of thermal and atmospheric scenarios~\citep{Alei2022,Konrad2022}.
Since the peak thermal emission in runaway greenhouse atmospheres will substantially decrease the star-to-planet flux ratio, mid-infrared wavelengths offer the possibility to probe the diversity of runaway climates in systems across different ages~\citep{2014ApJ...784...27L,2019A&A...621A.125B}.
Finally, mid-infrared surveys such as \life\ show a preference for M~star planets~\citep{Quanz2022}, which is beneficial for detecting the predicted habitable zone inner edge discontinuity (see Section~\ref{sec:results_FGK_M}).
With a sample size of a few tens of planets crossing the runaway greenhouse transition, direct imaging missions will thus enable key insights into the compositional inventory of atmospheric volatiles and climate states~\citep{2021exbi.book....5H,2022arXiv220505696C}, adding important details to a potential runaway greenhouse detection purely via transit radii.

As for exoplanets missions in their implementation phase, however, \plato\ overall remains to be the most promising mission for an empirical confirmation or falsification of the runaway greenhouse transition at this time.

\subsection{Mission design trade studies}\label{sec:mission-design-trades}
To explore the impact of mission trades on the detectability of the habitable zone inner edge discontinuity, we simulated different survey designs and strategies and measured their capability to recover the trend and constrain its parameters.
We assessed this capability based on two determinants: the likelihood that the mission is able to detect the injected trend, and the precision with which it can constrain the parameters of that trend.

\subsubsection{The value of follow-up campaigns}
The constraining power changes when additional information beyond planet radii is available for the characterized planet population.
As runaway greenhouse phases leave a stronger imprint on bulk density than on planet radius (see Section~\ref{sec:res_followup}), it would be beneficial to obtain constraints on planetary masses and test the runaway greenhouse hypothesis in density space instead of radius space.
This way, useful results can be obtained under more pessimistic conditions, e.g., a low predominant water content of planetary surfaces and atmospheres or a smaller available planet sample.
For a mission design similar to \plato\, a density-based hypothesis test on about a third of the overall sample is equivalent to a pure radius-based analysis.
At a fixed sample size, key parameters of the runaway greenhouse models can be more narrowly constrained when additional mass measurements are available.

Precise ground-based radial velocity measurements will be needed to provide these data, and a number of instruments are already successfully employed in characterizing terrestrial-sized exoplanets~\citep[e.g.,][]{Queloz2001a,Pepe2010,Johnson2010b,Ribas2023} and confronting these results with planet formation theory~\citep[e.g.,][]{Miguel2020a,Burn2021,Zawadzki2021a,Schlecker2022}.
A new generation of instruments on extremely large telescopes such as \gclef\ on the \gmt ~\citep{Szentgyorgyi2016}, \andes\ on the \elt ~\citep{Marcantonio2022}, or \modhis\ on the \tmt ~\citep{Mawet2019} will open up the discovery space even further.

Recently, NASA and the National Science Foundation (NSF) commissioned an ``Extreme Precision Radial Velocity Initiative''~\citep{Crass2021} to develop methods and facilities for precise mass measurements of temperate terrestrial planets.
Their findings highlight that such measurements are costly, and therefore follow-up efforts may only be available for a subsample of the targets of a mission of \plato's scale.
The diagnostic power of the hypothesis tests we demonstrated here may be improved by simultaneously fitting for the habitable zone inner edge discontinuity in the subsample without RV follow-up.
An optimized mission in search for the inner edge of the habitable zone will further enhance its information content via an informed selection of follow-up targets, i.e., balancing objects located on either side of the expected instellation threshold $S_\mathrm{thresh}$.

\subsubsection{The importance of M~dwarfs in the target list}
To date, the majority of planets with radius measurements orbit FGK~dwarfs, and, based on the instellation they receive, most of them lie in the runaway greenhouse regime~\citep{Thompson2018}.
Obviously, a radius/density discontinuity in the exoplanet demographics like the habitable zone inner edge discontinuity cannot be constrained well if only one side of the discontinuity is being sampled.
This, however, is the situation for planetary systems around Sun-like stars -- their habitable zones are so distant that transiting planets within them are very rare due to pure geometrical reasons.
It was, among other reasons, the sharp drop in transit probability with orbital distance that has prompted a number of recent transit surveys to specifically target M~dwarfs~\citep[e.g.,][]{Irwin2009,Obermeier2016,Delrez2018,Sebastian2021,Dietrich2023}, but the sample of terrestrial planets orbiting them is still small~\citep[e.g.,][]{Berger2020,Hardegree-Ullman2020a}.

M~dwarf systems are also key for detecting the runaway greenhouse transition:
Our calculations with different spectral types (Section~\ref{sec:results_FGK_M}) show that the information content of M~dwarfs in a sample dominates the hypothesis tests.
Besides their large number in a volume- or magnitude-limited sample, transiting M~dwarf planets are more likely to be located near the threshold instellation and in particular on orbits further out, i.e., in the optimistic habitable zone.
In fact, we showed that the FGK part of the planet sample barely contributes to the statistical power.
Furthermore, the transit depth difference at the transition is expected to be larger for M~dwarfs~\citep[$\sim \SI{100}{\ppm}$ for early, $\sim \SI{1000}{\ppm}$ for late M~stars,][]{Turbet2019}, enhancing the demographic signal it leaves.
An additional advantage of targeting M~dwarfs is \rev{the extended runaway greenhouse phases of their planets that can last on} the order of gigayears~\citep{Luger2015}.
This increases the probability of observing any given planet in the sample during the runaway greenhouse phase, essentially driving $f_\mathrm{rgh}$ to higher values.
Therefore, in addition to the high scientific value of boosted detections of potentially habitable planets, M dwarfs are also indispensable for the discovery and characterization of the runaway greenhouse transition.
As with a pure volume-limited sample, a targeted M~dwarf survey, too, profits from follow-up measurements of planetary masses with an order of magnitude increase in evidence.

\subsection{Constraining planetary habitability}\label{sec:habitability}
A potential for liquid water on the surface of a planet is commonly used as an environmental marker to assess its surface habitability~\citep{Huang1959,Hart1978,Kasting1993,Kaltenegger2011,Kopparapu2013}.
The runaway greenhouse transition represents an upper bound on received irradiation for this condition.
Its detection would thus not only empirically confirm the habitable zone concept but also help to locate it in the observationally available planetary parameter space.
In Section~\ref{sec:res_constrain-S_thresh}, we show that the threshold instellation at which the runaway greenhouse transition occurs can be reasonably constrained without imposing overly optimistic conditions on the underlying planet population, instrumentation, or survey strategy.
A mission like \plato\ is well equipped to perform this measurement; the constraining power is directly proportional to the proportion of characterized planets around M dwarfs and to the number of planets for which masses can be determined.

The situation is different for the planetary water inventory and the fraction of planets with runaway greenhouse climates:
Since these parameters are degenerate, they cannot be well constrained without independent measurements.
This degeneracy could be lifted if independent measurements of atmospheric compositions can be made.
For example, detections of water vapor in planets above the threshold instellation, combined with precise radius measurements, would constrain the predominant water content of terrestrial planets.

Once a runaway greenhouse region is identified in the parameter space, the community will  have a tool at hand to discern potentially habitable planets from Venusian worlds on an empirical basis.
Together with atmospheric measurements (see Section~\ref{sec:dis_atmospheres}), we will be able to put a number on the probability of an individual planet to harbor sufficient surface water to sustain life.

\subsection{Impact of assumptions on our findings}
The prospects for probing the runaway greenhouse transition depend on astro- and geophysical factors, as well as on the specific instrumentation and survey strategy of a particular mission.
Our state-of-the-art models approximate the situation and offer testable predictions.
In the following, we review a few considerations that future models may include to refine these predictions.

\subsubsection{Structures in the planet occurrence rate density}
The baseline occurrence rate density in radius-period space that governs the generation of synthetic planets might influence our findings, especially if it contains any features that coincide with the injected demographic feature.
This is not the case in the model from \citet{Bergsten2022} that we adopted: Its occurrence rate density varies smoothly in the domain relevant for the runaway greenhouse hypothesis; transitions only occur at smaller instellations ($< \SI{50}{\watt\per\meter\squared}$) and larger radii ($> 1.6\,R_\oplus$).
We thus do not expect the model underlying our planet sample to affect the investigation of the habitable zone inner edge discontinuity.
If any currently unknown sharp features in the distribution of terrestrial planets emerge, they should be considered in future studies.

\subsubsection{Baseline mass–radius relationship}
Our baseline mass–radius relationship assuming pure $\mathrm{MgSiO_3}$ interiors~\citep{Zeng2016} might not be representative of the rocky planet population.
However, while interior composition may introduce an offset to the radius habitable zone inner edge discontinuity, we do not expect a change of its structure.
Since the magnitude of the radius inflation effect is expected to be larger for an Earth-like interior composition with an iron core-silicate mantle structure~\citep{Zeng2016,Noack2020,2021JGRE..12606724B}, we consider our mass–radius relationship a conservative case.
We performed a sanity check to assess the impact of varying our baseline model (see Appendix~\ref{app:MR_relation}) and found general agreement between different interior compositions.

Future self-consistent modeling of interior-atmosphere interactions may include constraints on additional radius increases due to a molten interior~\citep{Bower2019} and any potential effects stemming from a deviating gas exchange between atmosphere and interior in runaway greenhouse planets due to different redox conditions~\citep{Ikoma2018,Lichtenberg2021c,2022PSJ.....3...93B,2021SSRv..217...22G} or water outgassing efficiency~\citep[e.g.,][]{Hier-Majumder2017,Ikoma2018,Salvador2023}.

\subsubsection{Bulk water mass fraction}
The predominant mass fractions of water, which sensitively controls the atmospheric state of a rocky exoplanet, is poorly constrained.
Inferred water contents in the literature range from upper limits on the order $10^{-5}$ to ``water worlds'' with tens of percent mass fraction~\citep[e.g.,][]{Rogers2010,Unterborn2018,Mousis2020,Agol2021,Luque2022}, all of which are within the realm of theoretical predictions\rev{~\citep{Selsis2007,Mulders2015b,Sato2016,Jin2018,Lichtenberg2019,Bitsch2019b,Venturini2020,Emsenhuber2021b,Schlecker2021,2022ApJ...938L...3L,2022ApJ...939L..19I}}.
Our nominal case assumes a bulk water mass fraction of $x_\mathrm{H_2O} = \var{wrr}$.
This can be considered a conservative choice that is unlikely to introduce a systematic overestimation of the habitable zone inner edge discontinuity. 
Cases of pure rocky composition and very low volatile contents can be considered absorbed by the dilution factor $f_\mathrm{rgh}$.
Assuming a distribution of water mass fractions instead of a fixed value would thus not significantly change our results.

\subsubsection{``Sharpness'' of the habitable zone inner edge discontinuity}
The habitable zone inner edge discontinuity may be affected by several processes that are challenging to quantify:
Planets that lack an atmosphere, sufficient volatiles, or have non-water-dominated outgassed compositions cannot bear steam atmospheres, and those that do eventually move to the non-runaway greenhouse category due to desiccation~\citep[][]{Watson1981,Kasting1983,Hamano2013} or evolution of their host star~\citep{Luger2015}.
Hydrogen/Helium-dominated planets may disguise as inflated rocky planets and not contribute to the demographic signal, although a runaway greenhouse radius inflation effect was suggested for water-dominated sub-Neptunes~\citep{Pierrehumbert2022,Innes2023}.
A subset of such gas-rich planets will experience atmospheric loss via photoevaporation~\citep{Owen2013} or core-powered mass loss~\citep{Ginzburg2018}, reducing their transit radius.
Intrinsic variation in the threshold instellation is caused by differences in planetary features influencing the onset of a runaway climate such as albedo, atmospheric composition, clouds, or surface gravity\reviii{~\citep{Salvador2017,Turbet2021,Lichtenberg2021c,Pierrehumbert2022,Turbet2023,Innes2023}}.
The choice of a statistical estimator for the hypothesis tests may further influence the recovered discontinuity; we compare our nominal running mean approach with a binned statistic in Appendix~\ref{app:binnedstats}.

While these factors may offset the signal's amplitude, they preserve its general shape.
The ``dilution factor'' $f_\mathrm{rgh}$ in our model embodies our ignorance of the magnitude of this offset.
In a real survey, additional contextual information about planets in the sample may be available.

\newpage[4]
\section{Conclusions}\label{sec:conclusions}
Significant inflation of rocky planet radii is a robust prediction of runaway greenhouse models.
Using \bioverse, a quantitative hypothesis testing framework, we have explored the potential of contemporary exoplanet missions to statistically detect a radius/density discontinuity resulting from this inflation in the exoplanet population.
Our key findings are as follows:
\begin{enumerate}
    \item The predicted runaway greenhouse transition causes a discontinuity in the radius and density distribution of small exoplanets with respect to their irradiation.
    \item This habitable zone inner edge discontinuity should be detectable with high-precision transit measurements.
          For a planet sample $\gtrsim 100$, a detection is likely if radius inflation occurs on at least \SI{10}{\percent} of the observed planets and if typical bulk water mass fractions are above $\sim 10^{-3}$.
    \item We find that the planned \plato\ transit survey will provide a sufficient sample and the required precision to confirm or reject the predicted trend.
          Assuming the projected photometric precision, \plato\ will be able to test the runaway greenhouse hypothesis for planet yields $\gtrsim 100$.
    \item The diagnostic power of transit missions in testing this hypothesis can be increased through a follow-up campaign providing planet mass measurements.
          This can reduce the required planet yield by about a factor of three.
          Only an adequate sample of planets orbiting M~dwarfs will ensure sufficient targets on both sides of the expected threshold instellation.
    \item Testing the runaway greenhouse hypothesis on a population level can provide constraints on the water inventory of rocky exoplanets and thus make an important contribution to assessing their habitability.
         A detection will provide an empirical confirmation of the habitable zone concept and localize its inner edge.

\end{enumerate}

The habitable zone concept is widely employed in target prioritization for exoplanet missions, and it will provide context for interpreting potential signatures of life.
As we have demonstrated, it appears realistic that an empirical test of the habitable zone hypothesis is imminent.
The confirmation or rejection of the habitable zone inner edge discontinuity will be a key contribution to understanding the diversity of exoplanet climates and the search for extraterrestrial life in the Universe.

\pagebreak[4]
\section*{Acknowledgments}
The authors thank Gabriele Cugno, Yann Fabel, Brad Foley, Lisa Kaltenegger, Ravi Kopparapu, Eric Mamajek, Megan Mansfield, Paul Molliere, Gijs Mulders, Matthew Murphy, Sukrit Ranjan, and Terry-Ann Suer for insightful discussions.
We are grateful to Martin Turbet for providing the mass–radius relationship data for planets harboring steam atmospheres.
We thank the anonymous referees for providing insightful reports that helped to improve this manuscript.
This material is based upon work supported by the National Aeronautics and Space Administration under agreement No. 80NSSC21K0593 for the program ``Alien Earths''.
The results reported herein benefited from collaborations and/or information exchange within NASA’s Nexus for Exoplanet System Science (NExSS) research coordination network sponsored by NASA’s Science Mission Directorate.
This work has made use of data from the European Space Agency (ESA) mission \gaia\ (\url{https://www.cosmos.esa.int/gaia}), processed by the \gaia\ Data Processing and Analysis Consortium (DPAC, \url{https://www.cosmos.esa.int/web/gaia/dpac/consortium}). Funding for the DPAC has been provided by national institutions, in particular the institutions participating in the \gaia\ Multilateral Agreement.
T.L. was supported by a grant from the Branco Weiss Foundation.
G.B. acknowledges support from the NASA Astrophysics Data Analysis Program under grant No.~80NSSC20K0446.
A.S. acknowledges support from NASA's Habitable Worlds Program (No.~80NSSC20K0226).

\section*{Author contributions}
M.S., D.A., and T.L.\ conceived the project, planned its implementation, and interpreted the results.
D.A.\ leads the ``Alien Earths'' program through which this project is funded and helped to guide the strategy of the project.
T.L.\ and A.S.\ provided expertise on runaway greenhouse climates and exoplanet interiors.
M.S.\ carried out the hypothesis tests and statistical analyses.
M.S.\ wrote the manuscript; T.L., G.B., K.H.-U., and A.S.\ provided text contributions.
G.B.\ implemented the planet generator in the \bioverse\ framework.
All authors provided comments and suggestions on the manuscript.

\section*{Reproducibility}
This study uses the reproducibility framework ``showyourwork''~\citep{Luger2021}.
All code required to reproduce our results, figures, and this article itself is available at \url{https://github.com/matiscke/hz-inner-edge-discontinuity}\rev{, and the repository state at the time of paper acceptance can be found at \dataset[doi:10.5281/zenodo.8251077]{\doi{10.5281/zenodo.8251077}}}.
The code to reproduce a figure can be accessed via the icon link next to the respective figure caption.
\rev{Data sets associated with this work are available at \dataset[doi:10.5281/zenodo.7080391]{\doi{10.5281/zenodo.7080391}}~(stellar luminosity tracks from \citealt{Baraffe1998}) and \dataset[doi:10.5281/zenodo.7946446]{\doi{10.5281/zenodo.7946446}} (results from model grid runs of \bioverse).}

\software{
\bioverse ~\citep{Bixel2021},
Astropy~\citep[][]{AstropyCollaboration2018},
NumPy~\citep[][]{Harris2020},
SciPy~\citep[][]{Virtanen2020},
corner.py~\citep{Foreman-Mackey2016b},
dynesty~\citep{Speagle2020}.
}

\appendix
\section{Robustness tests}
\subsection{Alternative statistics for the average radius or bulk density}\label{app:binnedstats}
The hypothesis tests introduced in Section~\ref{sec:met_surveys-hypotests} rely on a statistical estimator for the variation of planetary radii or bulk densities as a function of net instellation, and we chose a moving average for this estimator in our nominal setup.
Here, we explore how robust our results are against this choice by demonstrating the recovery of the runaway greenhouse signal in the case of the optimistic survey (Section~\ref{sec:res_testability}) with an alternative estimator: instead of computing moving averages, we used a binned statistic.

We first binned the data of the simulated survey in instellation space, choosing the number of bins via the rule of \citet{Freedman1981} and using logarithmic binning.
In each bin, we computed the arithmetic mean of the planet radius and its standard deviation.
Then, we assigned each planet the mean radius according to the instellation bin it occupies and used this as the measure for testing the runaway greenhouse hypothesis.

\begin{figure}[ht!]
    \script{optimistic_RS_binned.py}
    \begin{centering}
        \includegraphics[width=\linewidth]{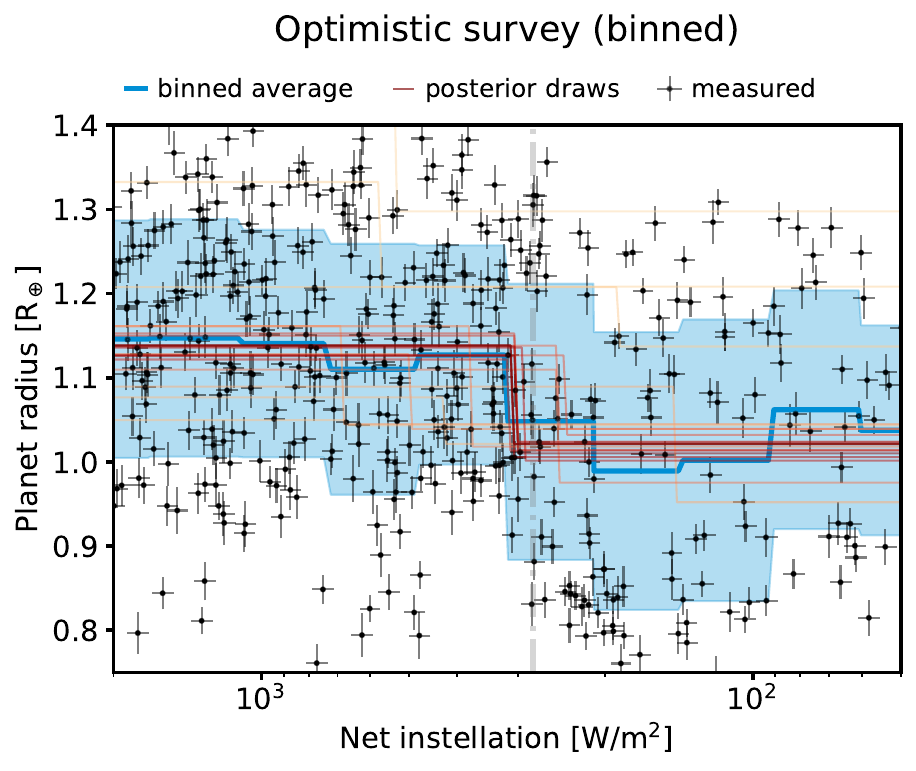}
        \caption{
            Detection of the runaway greenhouse threshold with a binned statistic.
            Using simulated data from the optimistic survey case ($N = \var{N_optimistic}$, compare Figure~\ref{fig:optimistic_R-S}), we tested the runaway greenhouse hypothesis. Instead of using a running mean, we computed a binned statistic (arithmetic mean and standard deviation in blue) to assign planets the average radius or bulk density in their neighborhood in instellation space. Random draws from the posterior of the runaway greenhouse hypothesis (Equation~\ref{eq:rgh_hypo}) are shown in red.
            As in the nominal case, the pattern is detected with high significance.
        }
        \label{fig:optimistic_R-S_binned}
    \end{centering}
\end{figure}
Figure~\ref{fig:optimistic_R-S_binned} shows the simulated data together with binned, average planet radii and draws from the posterior of the hypothesis test.
A clear detection resulted, although with somewhat lower significance ($\Delta \ln \mathcal{Z} \approx 30$) compared to the nominal setup.
The accuracy of the recovered instellation threshold is comparable.
This test demonstrates that our results are not sensitive to the choice of statistical estimator to test the hypotheses against.

It is conceivable that with very large sample sizes and a very sharp runaway greenhouse transition a binned solution would perform better.
For a search with real data, both approaches, and possibly other alternatives, should be considered.

\subsection{Statistical imprint of runaway greenhouse atmospheres}\label{app:model-pop_comparison}
The predicted runaway greenhouse-induced planet radius changes are a function of instellation, planet mass, and \rev{bulk} water mass fraction~(compare Section~\ref{sec:met_rghmodel}).
In order to better understand the interaction between the planetary populations underlying our simulations and these predictions, we compared the latter to the average radius and bulk density changes we measured in the synthetic population.
We used the ``optimistic'' scenario with a sample size of \var{N_optimistic}.

\begin{figure*}[ht!]
    \script{model_pop_comparison.py}
    \begin{centering}
        \includegraphics[width=\linewidth]{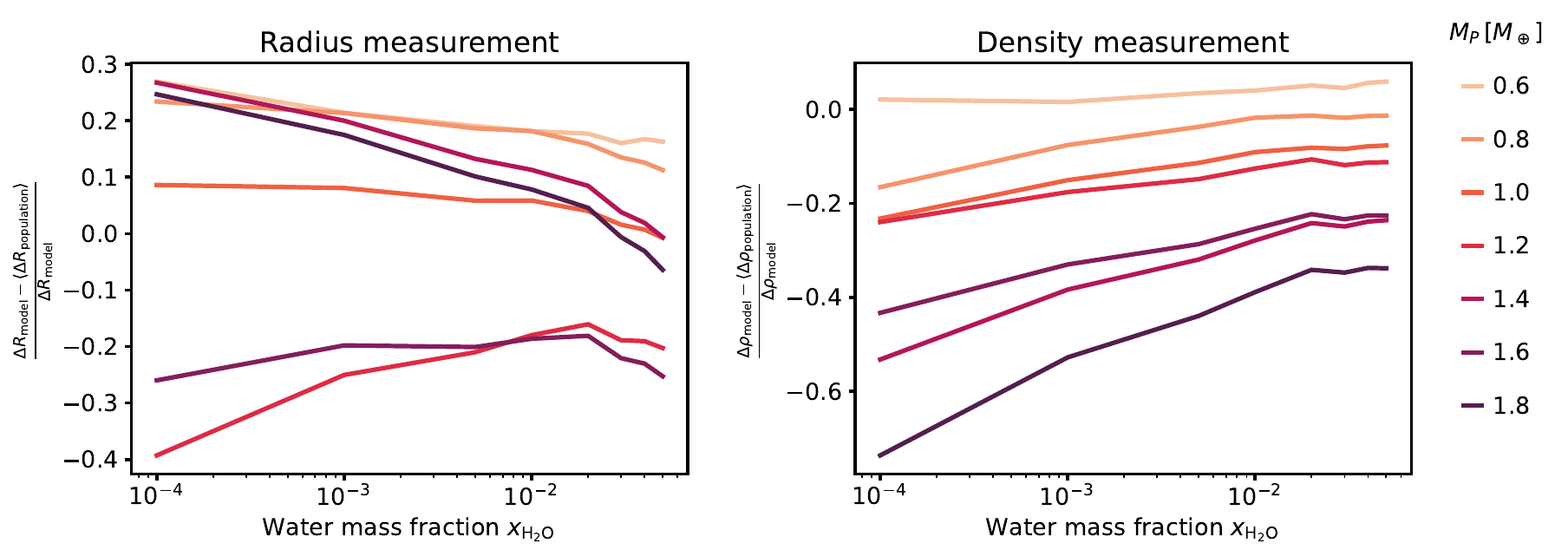}
        \caption{
        Comparison of radius and bulk density changes predicted by the atmospheric models to the average changes measured in the synthetic planet population.
        Model predictions $\Delta R_\mathrm{model}$ and $\Delta \rho_\mathrm{model}$ depend on individual planet masses; $\langle \Delta R_\mathrm{population} \rangle$ and $\langle \Delta \rho_\mathrm{population} \rangle$ are averaged measurements of the overall population.
        Significant differences are thus expected.
        }
        \label{fig:model_pop_comparison.pdf}
    \end{centering}
\end{figure*}
Figure~\ref{fig:model_pop_comparison.pdf} shows this comparison for a range of planetary masses and water mass fractions.
The complex dependence of the radius inflation on these parameters is evident, but no significant abrupt changes capable of causing spurious signals occur.
Differences between model prediction and population can be explained by the wide dispersion in planet mass in the population.

\subsection{Influence of different mass–radius relationships on our results}\label{app:MR_relation}
We assessed how a different choice of baseline mass–radius relation influences our results.
Focusing on the detectability of the statistical runaway greenhouse signal and the ability to constrain the threshold instellation, we repeated the hypothesis test in Section~\ref{sec:res_testability} with alternative mass–radius relationships.
Instead of assuming a pure $\mathrm{MgSiO_3}$ composition, we assigned planet masses using either the probabilistic relationship in \citet{Wolfgang2016} or a semi-empirical, two-layer relation assuming an Earth-like (\SI{32.5}{\percent} Fe + \SI{67.5}{\percent} MgSiO$_3$) composition~\citep{Zeng2016}.

\begin{figure}[ht!]
    \script{MR_violins.py}
    \begin{centering}
        \includegraphics[width=\linewidth]{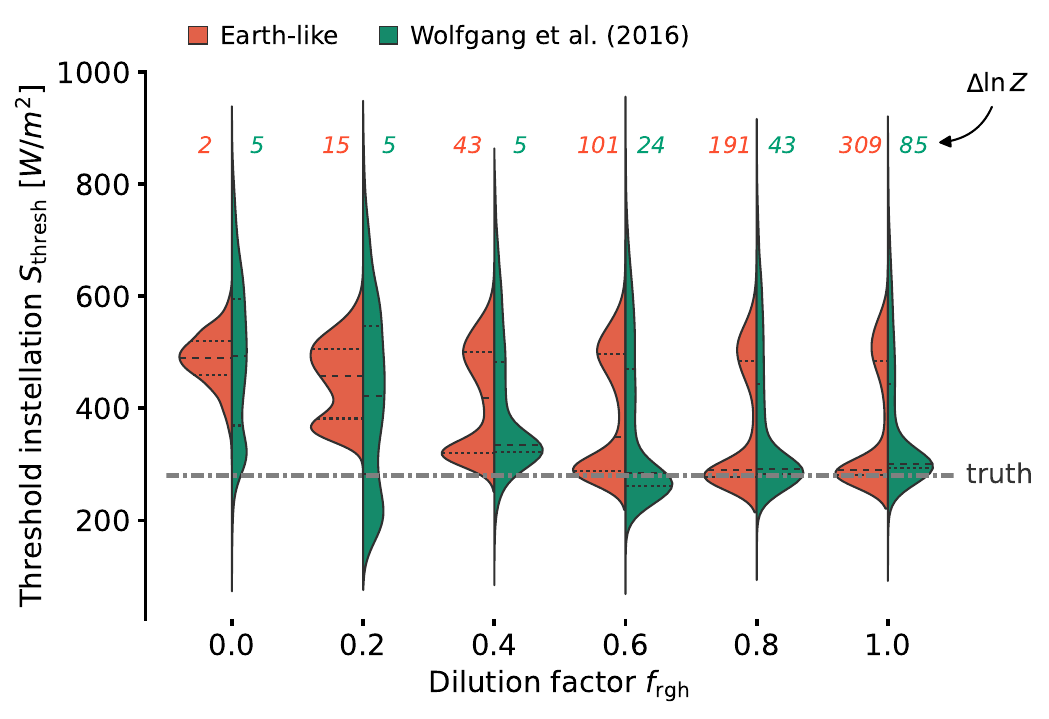}
        \caption{
        Posterior threshold instellations as a function of dilution factor for two alternative baseline mass–radius relations.
        For each grid step in $f_\mathrm{rgh}$, we show kernel density estimates of retrieved posteriors (averaged over 50 iterations) assuming an Earth-like (\SI{32.5}{\percent} Fe + \SI{67.5}{\percent} MgSiO$_3$) composition or the probabilistic relation from \citet{Wolfgang2016} and otherwise following the optimistic scenario in Section~\ref{sec:res_testability}.
        Lines within the violins show quartiles of the distributions, and the gray line indicates the injected threshold instellation of \var{injectedSthresh}.
        Higher log-evidence differences $\Delta \ln Z$ correspond to more significant rejections of the null hypothesis.
        In the regime of strong detections, both mass–radius relations lead to similar, accurate constraints on $S_\mathrm{thresh}$.
        Differences occur at low dilution factors, where the Earth-like relation leads to narrower estimates.
        }
        \label{fig:MR-violins}
    \end{centering}
\end{figure}
Figure~\ref{fig:MR-violins} shows how the two alternative baseline mass–radius relations influence the significance of a detection and the ability to constrain the threshold instellation.
The relation of \citet{Wolfgang2016} includes intrinsic scatter, which impedes a detection at low dilution factors.
In this regime, a layered, Earth-like composition leads to more significant detections and a narrower, although biased, constraint on $S_\mathrm{thresh}$.
Both mass–radius relations agree and recover the injected value where the null hypothesis can be rejected with high significance.
This is consistent with our nominal mass–radius relation (compare Section~\ref{sec:met-orbits_masses}).
We conclude that the underlying core and mantle composition of planets may affect the detectability of the transition if the fraction of planets with runaway greenhouse climates is low; however, the overall trends that our experiments revealed appear robust.

We caution that this analysis may serve only as a sanity check and should not be taken as a result in itself: The atmospheric model from \citet{Turbet2020} we adopted relies on a silicate interior composition for its transit radius prediction.
Therefore, only our nominal procedure throughout the main body of the paper represents a self-consistent treatment.

\bibliographystyle{aasjournal}
\bibliography{bib,PhD,coauthors,refs_tim}

\end{document}